\documentclass[useAMS,usenatbib,twocolumn]{mnras}
\usepackage{natbib}
\usepackage{graphicx}
\usepackage{rotate}
\usepackage{hyperref}
\usepackage{breakurl}

\newcommand{\eg}{{\it e.g.,}}

\newcommand{\kms}{\ensuremath{\mathrm{km\,s}^{-1}}}

\newcommand{\Mpc}{\ensuremath{\mathrm{Mpc}}}

\newcommand{\ds}{\ensuremath{\Delta\Sigma}}
\newcommand{\photoz}{photo-$z$~}
\newcommand{\photozs}{photo-$z$s~}

\newcommand{\h}{\ensuremath{h^{-1}}~}
\def\lcdm{\Lambda{\rm CDM}}

\newcommand{\textcol}{\textcolor{black}{}}

\begin{document}
  

\title[Lensing of CMASS]{Lensing is Low: Cosmology, Galaxy Formation, or New Physics?}




 \author[Leauthaud et al.]  {Alexie Leauthaud$^{1,2}$, Shun Saito$^3$, Stefan Hilbert$^{4,5}$, Alexandre Barreira$^3$, Surhud More$^2$,\newauthor
  Martin White$^6$, Shadab Alam$^{7,8}$, Peter Behroozi$^{6,9}$, Kevin Bundy$^{1,2}$, Jean Coupon$^{10}$,\newauthor
  Thomas Erben$^{11}$, Catherine Heymans$^{8}$, Hendrik Hildebrandt$^{11}$, Rachel Mandelbaum$^{7}$,\newauthor
  Lance Miller$^{12}$, Bruno Moraes$^{13}$, Maria E. S. Pereira$^{14}$, Sergio A. Rodr\'iguez-Torres$^{15,16,17}$,\newauthor
 Fabian Schmidt$^3$, Huan-Yuan Shan$^{18}$, Matteo Viel$^{19,20}$, Francisco Villaescusa-Navarro$^{19,20,21}$\\
$^1$Department of Astronomy and Astrophysics, University of California, Santa Cruz, 1156 High Street, Santa Cruz, CA 95064 USA\\
$^2$Kavli IPMU (WPI), UTIAS, The University of Tokyo, Kashiwa, Chiba, 277-8583, Japan\\
$^3$Max-Planck-Institut f\"{u}r Astrophysik, Karl-Schwarzschild-Star{\ss}e 1, D-85740 Garching bei M\"{u}nchen, Germany\\
$^4$Exzellenzcluster Universe, Boltzmannstr. 2, 85748 Garching, Germany\\   
$^5$Ludwig-Maximilians-Universit{\"a}t, Universit{\"a}ts-Sternwarte, Scheinerstr. 1, 81679 M{\"u}nchen, Germany\\  
$^6$Department of Physics, University of California, Berkeley, CA 94720\\
$^7$McWilliams Center for Cosmology, Department of Physics, Carnegie Mellon University, Pittsburgh, PA 15213\\
$^{8}$The Scottish Universities Physics Alliance, Institute for Astronomy, University of Edinburgh, Blackford Hill, Edinburgh EH9 3HJ, UK\\
$^9$Hubble Fellow\\
$^{10}$Astronomical Observatory of the University of Geneva, ch. d'Ecogia 16, 1290 Versoix, Switzerland\\
$^{11}$Argelander-Institut f\"ur Astronomie, Auf dem H\"ugel 71, D-53121 Bonn, Germany\\
$^{12}$Dept. of Physics, University of Oxford, Oxford OX1 3RH, UK\\
$^{13}$Dept. of Physics and Astronomy, University College London, London, WC1E 6BT, UK\\
$^{14}$Centro Brasileiro de Pesquisas Fisicas - Rua Dr. Xavier Sigaud 150, CEP 22290-180, Rio de Janeiro, RJ, Brazil\\
$^{15}$Instituto de F\'isica Te\'orica, (UAM/CSIC), Universidad Aut\'onoma de Madrid, Cantoblanco, E-28049 Madrid, Spain\\
$^{16}$Campus of International Excellence UAM+CSIC, Cantoblanco, E-28049 Madrid, Spain\\
$^{17}$Departamento de F\'isica Te\'orica M8, Universidad Aut\'onoma de Madrid (UAM), Cantoblanco, E-28049, Madrid, Spain\\
$^{18}$Laboratoire d'astrophysique (LASTRO), Ecole Polytechnique F\'ed\'erale de Lausanne (EPFL), Observatoire de Sauverny, CH-1290 Versoix, Switzerland\\
$^{19}$INAF, Osservatorio Astronomico di Trieste, via Tiepolo 11, I-34131 Trieste, Italy\\
$^{20}$INFN -- National Institute for Nuclear Physics, Via Valerio 2, I-34127 Trieste, Italy\\
$^{21}$Center for Computational Astrophysics, 160 5th Ave, New York, NY, 10010, USA}

\maketitle
\label{firstpage}

 
\begin{abstract} We present high signal-to-noise galaxy-galaxy lensing measurements of the BOSS CMASS sample using 250 square degrees of weak lensing data from CFHTLenS and CS82. We compare this signal with predictions from mock catalogs trained to match observables including the stellar mass function and the projected and two dimensional clustering of CMASS. We show that the clustering of CMASS, together with standard models of the galaxy-halo connection, robustly predicts a lensing signal that is 20-40\% larger than observed. Detailed tests show that our results are robust to a variety of systematic effects. Lowering the value of $S_{\rm 8}=\sigma_{\rm 8} \sqrt{\Omega_{\rm m}/0.3}$ compared to \citet{Planck-Collaboration:2015aa} reconciles the lensing with clustering. However, given the scale of our measurement ($r<10$ \h Mpc), other effects may also be at play and need to be taken into consideration. We explore the impact of baryon physics, assembly bias, massive neutrinos, and modifications to general relativity on $\Delta\Sigma$ and show that several of these effects may be non-negligible given the precision of our measurement. Disentangling cosmological effects from the details of the galaxy-halo connection, the effects of baryons, and massive neutrinos, is the next challenge facing joint lensing and clustering analyses. This is especially true in the context of large galaxy samples from Baryon Acoustic Oscillation surveys with precise measurements but complex selection functions. \end{abstract}

 

 
 


\section{Introduction}

Weak gravitational lensing is recognized as a powerful and unique cosmological tool because it is one of the few direct probes of the total mass distribution of the universe, including the dark matter component. The weak lensing signal around galaxy ensembles, commonly referred to as galaxy-galaxy lensing (hereafter ``g-g lensing"), provides a measure of the radial distribution of total mass around galaxies. Since its first detection two decades ago by \citet{Brainerd:1996}, g-g lensing has matured from a low signal-to-noise ($S/N$) novelty into a sophisticated cosmological probe with recent measurements reaching out to scales beyond 50 \h Mpc with high signal-to-noise \citep[\eg][]{Mandelbaum:2013}. With new large lensing surveys such as the Dark Energy Survey \citep[DES,][]{Dark-Energy-Survey-Collaboration:2016} and the Hyper Suprime Cam (HSC) survey\footnote{\url{http://hsc.mtk.nao.ac.jp/ssp/}} collecting high quality lensing data for thousands of square degrees, g-g lensing measurements will soon reach even greater precision. Space-based lensing missions such as \textit{Euclid} \citep{Laureijs:2011} and the Wide Field Infrared Survey Telescope \citep[WFIRST,][]{Spergel:2013} will launch within the next decade with even greater capabilities and in tandem, the Large Synoptic Survey Telescope \citep[LSST,][]{LSST-Science-Collaboration:2009} will collect 20,000 deg$^2$ of lensing quality data.

In parallel to these efforts, surveys such as the Baryon Oscillation Spectroscopic Survey \citep[BOSS,][]{Eisenstein:2011, Dawson:2013}, have collected optical spectra for more than one million massive galaxies at $z<1$. Within only a few years, next generation experiments such as the Dark Energy Spectroscopic Instrument \citep[DESI,][]{Levi:2013} and the Prime Focus Spectrograph \citep[PFS,][]{Takada:2014} will measure the redshifts of tens of millions of galaxies. Baryon Acoustic Oscillation (BAO) surveys yield exquisite measurements of galaxy clustering but also provide excellent samples for lensing studies because g-g lensing measurements are more robust when applied to lens samples with spectroscopic redshifts.

In addition, galaxy clustering and g-g lensing are large-scale structure probes with highly complimentary capabilities: the first measures the autocorrelation of galaxies whereas the second ties galaxies to the underlying dark matter distribution. Joint analyses that take advantage of the synergies between both probes are increasingly popular for studies of the galaxy halo connection \citep[\eg][]{Mandelbaum:2006c,Leauthaud:2012a,Coupon:2015,Zu:2015}, to constrain cosmological parameters \citep[\eg][]{Yoo:2006aa, Cacciato:2009, Cacciato:2013, More:2013, Mandelbaum:2013}, and to perform tests of General Relativity \citep[\eg][]{Reyes:2010,Blake:2016}.

Generally speaking, measurements of galaxy clustering and g-g lensing on non-linear scales ($r<$1 \h Mpc) provide us with detailed information about the galaxy-halo connection whereas large scale measurements ($r>$10 \h Mpc) are preferred for robust cosmological constraints because they can be modeled with linear theory and are less sensitive to galaxy formation processes. However, previous work has debated about exactly where to draw the line with some studies opting to remove small scale information entirely  at the cost of larger errors \citep[][]{Mandelbaum:2013}, whereas other work has included small scale information by marginalizing over the galaxy-halo connection \citep[][]{Cacciato:2013, More:2013, More:2015}. This connection is typically modeled either via Halo Occupation Distribution \citep[HOD, ][]{Jing:1998,Peacock:2000,Berlind:2002,Zheng:2005,Leauthaud:2011,Hearin:2016} or SubHalo Abundance Matching \citep[SHAM,][]{Kravtsov:2004,Conroy:2006,Behroozi:2010,Reddick:2013} type formalisms.

Until present, because of relatively modest lensing data sets, g-g lensing measurements have played more of an ancillary role compared to clustering measurements. However, with new rapidly expanding lensing surveys, g-g lensing is poised to play an increasingly important role in analyzing $z<1$ BAO samples. In particular, Redshift Space Distortions (RSD) and g-g lensing have important synergies for constraining the growth of structure. Unlike BAO measurements, RSD methods push into the semi non-linear regime and hence are more subject to theoretical systematics associated with the complexities of galaxy bias and need to be validated against realistic galaxy mock catalogs \citep{Alam:2016a}. With this in mind, g-g lensing measurements on both small and large scales will be important for cosmological constraints, but also for characterizing the details of the galaxy-halo connection to help pin down theoretical systematic uncertainties for RSD. This is especially true in the context of BAO samples which have complex selection functions and which are therefore non trivial to model using standard galaxy-halo type models.

This paper presents a high signal-to-noise ($S/N=30$) g-g lensing measurements for the BOSS ``constant mass" (CMASS) sample using 250 degrees$^2$ of lensing data (Sections  \ref{data} and \ref{ds_computation}). We show that the amplitude of the lensing signal is in tension with predictions from a variety of BOSS mock catalogs that reproduce the clustering of CMASS (Section \ref{results}). This may indicate that our data prefer a low value of the amplitude of matter fluctuations at low redshifts, a failure of standard models of the galaxy-halo connection, or may be a signature of the effect of baryons on the matter distribution. A discussion of our results, including detailed tests for systematic effects, is presented in Section {\ref{discussion} and summarized in Section \ref{conclusions}. We assume a flat $\Lambda$CDM cosmology with $\Omega_{\rm m}=0.31$, $\Omega_\Lambda=0.69$, H$_0=$100~h$^{-1}$~km~s$^{-1}$~Mpc$^{-1}$. Unless noted otherwise, distances are expressed in comoving coordinates.



\section{Data}\label{data}

\subsection{The BOSS CMASS Sample}\label{bosssample}

BOSS is a spectroscopic survey of 1.5 million galaxies over 10,000
deg$^2$ that was conducted as part of the SDSS-III program \citep[][]{Eisenstein:2011} on the 2.5 m aperture Sloan Foundation
Telescope at Apache Point Observatory \citep[][]{Gunn:1998,
  Gunn:2006}. A general overview of the BOSS survey can be found in
\citet[]{Dawson:2013} and the BOSS pipeline is described in
\citet[]{Bolton:2012}. BOSS galaxies were selected from Data Release 8
\citep[DR8,][]{Aihara:2011} {\it ugriz} imaging
\citep[][]{Fukugita:1996} using a series of color-magnitude cuts \citep[][]{Reid:2016}. BOSS
targeted two primary galaxy samples: the LOWZ sample at $0.15<z<0.43$
and the CMASS sample at $0.43<z<0.7$. In this paper, we focus on the
high redshift CMASS sample. As our input catalog, we use the BOSS DR11
\citep[][]{Ahn:2014} large-scale structure (LSS) catalog created by
the BOSS galaxy clustering working group \citep[][]{Anderson:2014}. 

Because each BOSS fiber has a diameter of 62\arcsec, no two objects
separated by less than this can be observed on a single plate. In
addition, redshift measurements fail for a small fraction of BOSS
galaxies ($<2$\% for CMASS). Because of these two effects, a small
number of CMASS targets do not obtain a spectroscopic redshift.

There are various different choices for how to correct for fiber
collisions and redshift failures -- details can be found in \citet{Anderson:2012} and \citet{Guo:2012aa}\footnote{Also see \url{http://www.sdss3.org/dr9/tutorials/lss_galaxy.php}.}. We test
several methods for dealing with galaxies with missing spectroscopic
redshifts and show that this correction does not have a large impact
on the CMASS lensing signal (see Appendix
\ref{app:fibercollisions}). As our fiducial correction method, we
adopt the same weighting scheme as the BOSS large-scale structure
working group. Namely, we upweight the galaxy nearest to each
unobserved galaxy (the ``nearest neighbor'' weighting method). Fiber
collision and redshift-failure correction weights are denoted $w_{\rm
  cp}$ and $w_{\rm rf}$, respectively.

Because our analysis is limited to relatively small scales, we do not
apply the angular systematic weights ($w_{\rm sys}$) or the minimum
variance weights ($w_{\rm FKP}$) that are used in BOSS large-scale
analyses (see Section 3 in \citealt{Anderson:2012} for details). Our
weighting scheme is consistent with the one adopted for the clustering
measurements of \citet[][]{Saito:2016}.

\subsection{Weak Lensing Data}

To measure the weak lensing signal of CMASS galaxies, we use a
combination of two data-sets: the Canada France Hawaii Telescope
Lensing Survey \citep[CFHTLenS,][]{Heymans:2012,Miller:2013} and the
Canada France Hawaii Telescope Stripe 82 Survey (CS82, Erben et al. in
prep). The combined area is $\sim$250 deg$^2$ and
both data sets use $i$'-band imaging from the CFHT MegaCam instrument
\citep[][]{Boulade:2003} taken under excellent seeing conditions
(seeing 0.6\arcsec-0.7\arcsec). Data reduction and shape measurements for
both surveys were performed homogeneously using the state-of-the-art
weak lensing pipeline developed by the CFHTLenS collaboration which
employs the \emph{lens}fit Bayesian shape measurement algorithm
\citep[][]{Heymans:2012,Miller:2013}. Differences between CFHTLenS and
CS82 that are of relevance for this work are the $i$'-band depth, the
source of additional photometry for \photoz measurements, and the
degree of overlap with the BOSS survey. Further details are now
described below.

\subsubsection{CFHTLenS Weak Lensing Catalog}

The CFHTLenS weak lensing catalogs are based on deep multicolor data
obtained as part of the CFHT Legacy Survey (CFHTLS). This survey spans
154 square degrees in five optical bands ($u^*$$g$'$r$'$i$'$z$') with
a 5$\sigma$ point source limiting magnitude of $i$'$\sim$25.5. Each
MegaCam pointing is roughly one square degree in area and has a pixel
size of 0.187 arcseconds. The CFHTLS Wide survey consists of four
separate patches on the sky known as W1, W2, W3 and W4 (63.8, 22.6, 44.2
and 23.3 deg$^2$ respectively). BOSS fully overlaps with the W4 field,
partially overlaps with W1 and W3, and only has a small amount of
overlap with W2. In this paper, we use the overlap regions in W1, W3,
and W4. Details on the image reduction, weak lensing shape
measurements, and photometric redshifts can be found in
\citet{Erben:2013}, \citet[]{Heymans:2012}, \citet[][]{Miller:2013},
and \citet[]{Hildebrandt:2012}.

We download the publicly available CFHTLenS weak lensing shear
catalogs\footnote{\url{http://www.cadc-ccda.hia-iha.nrc-cnrc.gc.ca/community/CFHTLens/query.html}}. Following
\citet[]{Heymans:2012}, we apply an additive calibration correction
factor, $c_2$, to the $\epsilon_2$ shape component on a galaxy-by-galaxy
basis\footnote{Equation 19 in \citet[]{Heymans:2012} assumes
  a galaxy size $r$ in arcseconds and the $r_0$ parameter has
  units of arcseconds.}. For each galaxy, we
also compute a multiplicative shear calibration factor as a function
of the signal-to-noise ratio and size of the source galaxy,
$m(\nu_{\rm SN},r)$\footnote{When using the parameters to
  compute $m$ provided in \citet[][]{Miller:2013}, $r$ corresponds to
  the {\sc scale-length} field in the CFHTLenS catalogs in pixel units.}. The calibration correction factor for $\epsilon_2$ and the multiplicative shear calibration factor, $m$, are computed separately for CFHTLenS and CS82. The values for the $m$ correction factor are given in Section \ref{combinedsignal} and represent a 3-7.5\% increase in $\Delta\Sigma$. Following \citet[][]{Velander:2014}, we do
not reject pointings that did not pass the requirements for cosmic
shear. The CFHTLenS {\it lens}fit catalogs contain a lensing (inverse
variance) weight $w$ which includes both the intrinsic shape
dispersion as well as the ellipticity measurement error. 

\subsubsection{CS82 Weak Lensing Catalog}

The Sloan Digital Sky Survey (SDSS) contains a sub-region of 275
deg$^2$ on the Celestial Equator in the Southern Galactic Cap known as
``Stripe 82'' \citep[][]{Abazajian:2009}. This region was repeatedly imaged during the Fall months when the North Galactic Cap was
not observable. The co-addition of these data reaches
$r\sim23.5$, about 2 magnitudes fainter than the SDSS single pass data
\citep[][]{Annis:2011} but has an $r$-band median seeing of
1.1\arcsec. 

The CS82 survey was designed with the goal of complementing existing
Stripe 82 SDSS {\it ugriz} imaging with high quality $i$'-band imaging
suitable for weak lensing measurements. CS82 is built from 173 MegaCam
$i$′-band images and corresponds to an area of 160 degrees$^2$ (129.2
degrees$^2$ after masking out bright stars and other image
artifacts). The Point Spread Function (PSF) for CS82 varies between
0.4\arcsec and 0.8\arcsec over the entire survey with a median seeing
of 0.6\arcsec. The limiting magnitude of the survey is
$i$′$\sim$24.1\footnote{The limiting magnitude is defined as the
  5$\sigma$ detection limit in a 2\arcsec~ aperture via
  $m_{\rm lim}=ZP-2.5\log(5)\sqrt{N_{\rm pix}\sigma_{\rm sky}}$, where $N_{\rm pix}$
  is the number of pixels in a circle of radius 2\arcsec,
  $\sigma_{\rm sky}$ is the sky background noise variation, and $ZP$ is
  the zero-point.}. 

Image processing is largely based on the procedures presented in
\citet[][]{Erben:2009} and shear catalogs were constructed using the
same weak lensing pipeline developed by the CFHTLenS collaboration
using the \emph{lens}fit Bayesian shape measurement method
\citep[][]{Miller:2013}. We compute $m$ and $c_2$ for each galaxy and
construct a source catalog for CS82 in the same fashion as for
CFHTLenS. The CS82 source galaxy density is 15.8 galaxies
arcmin$^{-2}$ and the effective weighted galaxy number density (see
equation 1 in \citealt{Heymans:2012}) is 12.3 galaxies
arcmin$^{-2}$. Note that these numbers do not yet include any
\photoz quality cuts. These are described in the
following section.

\subsubsection{Photometric Redshifts}\label{photoz}

Photometric redshifts for the CFHTLenS source catalog have been
calculated by \citet[][]{Hildebrandt:2012} using the Bayesian
photometric redshift software {\sc bpz} \citep[][]{Benitez:2000,
  Coe:2006}. Photo-$z$s for CS82 have been calculated by
\citet[][]{Bundy:2015}, also using {\sc bpz}. For a redshift estimate, we
use $Z_B$, the peak of the posterior distribution given by {\sc
  bpz}. In addition to $Z_B$, we will use the 95 per cent
confidence limit (noted $\sigma_{95}$) on $Z_B$ as well as the {\sc
  bpz} {\sc odds} parameter (a measure of the peakiness of the
probability density function provided by {\sc bpz}). The {\sc odds}
parameter varies between 0 and 1 and galaxy samples with larger {\sc
  odds} values have reduced catastrophic outlier fractions
\citep[\eg][]{Margoniner:2008aa}.

Determining the level of systematic error due to photometric redshifts
is often one of the most uncertain aspects of a galaxy-galaxy weak
lensing analysis \citep[\eg][]{Nakajima:2012aa}. Fortunately, the CS82
survey overlaps with a number of existing spectroscopic surveys which
can be used to assess the quality of our photometric redshifts. We
compile a set of high quality spectroscopic redshifts that overlap
with CS82 from the Baryon Oscillation Spectroscopic Survey DR12 data
release \citep[BOSS;][]{Alam:2015}, VVDS
\citep[][]{Le-Fevre:2004}, DEEP2
\citep[][]{Newman:2013a}, and PRIMUS \citep[][]{Coil:2011}. For VVDS,
DEEP2, and PRIMUS, we select galaxies with a redshift quality flag
greater than or equal to 3. Our combined spectroscopic sample contains
a total of 11694 objects\footnote{207 galaxies from BOSS, 5328 from
  DEEP2, 4942 from PRIMUS, and 1217 from VVDS.}. Among these
data-sets, the DEEP2 redshifts are the most useful for our
purpose. The DEEP2 spectroscopic redshift catalog contains galaxies to
$R_{\rm AB}=24.1$ which spans the full magnitude range of our source
sample. However, the DEEP2 sample is also color selected to target
galaxies at $z>0.7$. Because we study lens galaxies between $z=0.43$
and $z=0.7$, a large majority of our source galaxies
have redshifts with $z>0.7$ which is well matched to the DEEP2
selection. Figure \ref{zphotzspec} displays a comparison between $z_{\rm spec}$
and $z_{\rm phot}$ for our fiducial CS82 source catalog (which
includes a cut of {\sc odds}$>0.5$).

When computing $\Delta\Sigma$, we do not
integrate over the full redshift Probability Distribution Function
(PDF), $p(z)$, of source galaxies. Indeed, \photoz codes
do not automatically provide accurate estimates for $p(z)$. For this reason, integrating over $p(z)$ does not automatically guarantee a more accurate
result. Instead, we use a point estimate as our source redshift, but we use an appropriately re-weighted version of the spectroscopic data set described above, to test for biases in our computation of $\Delta\Sigma$. The details of this computation are given in \ref{appendix_photoz}. Using
our combined set of spectroscopic redshifts, we estimate that \photoz errors cause our $\Delta\Sigma$ values to be over-estimated by $\sim$3\%. This estimate includes the dilution of the signal by source galaxies which have $z_{\rm spec}<z_L$ but $z_{\rm phot}>z_L$ where $z_L$ is the lens redshift. 
 
\begin{figure*}
\begin{center}
\includegraphics[width=18cm]{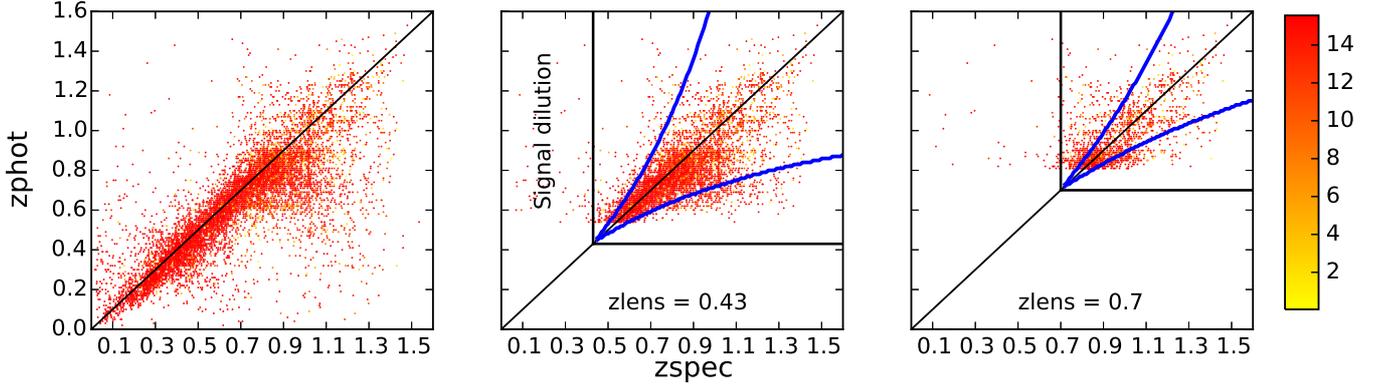}
\caption{Comparison between $z_{\rm spec}$ and $z_{\rm phot}$ for the
  CS82 source catalog. Galaxies are color coded according to their
  lensing weight $w$ (with values indicated by the color bar on the
  right hand side). Our CMASS lens sample is located at redshifts
  $0.43<z_L<0.7$. The middle and right hand panels display a comparison
  between $z_{\rm spec}$ and $z_{\rm phot}$ for ``background'' source
  galaxies for lens redshifts of $z_L=0.43$ and $z_L=0.7$. To select
  background galaxies, we require that $z_{S}>z_{L}+0.1$ and
  $z_{S}>z_{L}+\sigma_{95}/2.0$. Source galaxies that satisfy this cut
  but which have a true redshift $z_{\rm spec}<z_{L}$ will dilute the lensing
  signal. These correspond to objects located to the left of the
  vertical solid lines (middle and right panels) and
  represent less than 3\% of our source sample for lenses at
  $z_L=0.7$. Our fiducial cut of {\sc odds}$>0.5$ reduces the number
  of galaxies located in this region. Solid blue curves represent the
  locus corresponding to a 30\% bias on $\Delta\Sigma$. While we do have some source galaxies located outside the cone formed by the blue curves, what is important for $\Delta\Sigma$ is that mean value of $\Sigma_{\rm crit}$ (averaged over the full source population) is un-biased (see section \ref{appendix_photoz} and equation \ref{ds_equation3}).}
\label{zphotzspec}
\end{center}
\end{figure*}

Finally, we also perform a series of tests to demonstrate that our
lensing signal is robust to a variety of different cuts on the
\photoz catalog. The results of these tests are presented
in Section \ref{appendix_photoz}. No statistically
significant systematic trends are found for any of the tests that we
have implemented.

\subsubsection{Source Catalog and Background Selection}\label{backgroundselection}

We construct a source catalog by applying the following cuts: {\sc mask}$\leq 1$,
{\sc fitclass}$=0$, $i'<24.7$, and $w>0$. Here, {\sc fitclass} is a
flag to remove stars but also to select galaxies with well-measured
shapes (see details in \citealt{Miller:2013}) and {\sc mask} is a
masking flag. In addition, we also require that
each source galaxy has a \photoz estimate and we apply a
fiducial \photoz quality cut of {\sc odds}$>0.5$. Our
lensing signals are robust to the choice of this {\sc odds} parameter
cut (see Section \ref{appendix_photoz}). After applying these cuts,
the CFHTLenS and CS82 source catalogs correspond respectively to
effective weighted galaxy number densities\footnote{Here we use
  $n_{\rm eff}$ as defined by Equation 1 in \citet[]{Heymans:2012}. An
  alternative definition of $n_{\rm eff}$ is given by Equation 9 in
  \citet{Chang:2013aa}.} of $n_{\rm eff}=10.8$ galaxies arcmin$^{-2}$
and $n_{\rm eff}=4.5$ galaxies arcmin$^{-2}$.

To minimize any dilution of our lensing signal due to \photoz
uncertainties, we perform background selection by requiring that
$z_{S}>z_{L}+0.1$ and $z_{S}>z_{L}+\sigma_{95}/2.0$ where $z_{L}$ is
the lens redshift, $z_{S}$ is the source redshift, and $\sigma_{95}$ is the 95 per cent confidence limit on the source redshift. This fiducial scheme for separating background
sources from lens galaxies will be referred to as {\sc zcut2}. In
Appendix \ref{appendix_photoz} we show that our lensing signals are
robust to the exact details of this cut, which suggests that our
lensing signal is not strongly affected by contamination from
source galaxies with $z_{\rm spec}<z_L$ but $z_{\rm phot}>z_L$. Our tests with spectroscopic redshifts in Appendix \ref{appendix_photoz} confirms and quantifies this statement.

We do not apply a correction factor to account for a dilution effect from source galaxies that are
actually physically associated with our lens sample (the so-called
``boost correction factor''). A detailed justification of this choice
is presented in Appendix \ref{appendix_boost}.


\section{Computation of $\Delta\Sigma$}\label{ds_computation}

\subsection{Stacking Procedure}

Our stacking procedure closely follows the methodology outlined in
\citet[][]{Leauthaud:2012a} and we refer to that work for in-depth
details. The primary difference with respect to
\citet[][]{Leauthaud:2012a} is that here we stack the g-g
lensing signal in comoving instead of physical coordinates. The
g-g lensing signal that we measure yields an estimate of the
mean {\em surface mass density contrast} profile:

\begin{equation}
  \Delta \Sigma(r)\equiv\overline{\Sigma}(< r)-\overline{\Sigma}(r).
\label{dsigma}
\end{equation}

Here, $\overline{\Sigma}(r)$ is the azimuthally averaged and projected
surface mass density at radius r and $\overline{\Sigma}(< r)$ is the
mean projected surface mass density within radius r
\citep[][]{Miralda-Escude:1991,Wilson:2001}. The relationship between the tangential shear, $\gamma_{\rm t}$, and $\Delta\Sigma$ is given by:

\begin{equation}\label{dsequation}
\Delta\Sigma=\gamma_{\rm t} \Sigma_\mathrm{crit},
\end{equation}

\noindent where $\Sigma_\mathrm{crit}$ is the {\em critical surface mass density} which in comoving coordinates is expressed by:

\begin{equation}\label{eq:sigmacrit}
\Sigma_\mathrm{crit} = \frac{c^2}{4\pi G} \frac{D_A(z_S)}{D_A(z_L) D_A(z_L,z_S)(1+z_L)^2},
\end{equation}

\noindent where $D_A(z_L)$ and $D_A(z_S)$ are angular diameter distances to the lens and
source and $D_A(z_L,z_S)$ is the angular diameter distance between the lens
and source. 

The \emph{lens}fit algorithm provides an inverse variance weight, $w$, which can be used to optimally weight shear measurements. For a given lens $i$ and a given source $j$, the inverse variance weight for $\Delta\Sigma$ can be derived from Equation \ref{dsequation} and is equal to $w_{\mathrm{ds},ij}=w_j \Sigma_{\mathrm{crit},ij}^{-2}$. We use $w_{\mathrm{ds}}$ to compute $\Delta\Sigma$ via a weighted sum over all lens-source pairs:

\begin{equation}
  \Delta\Sigma =  
  {\sum_{i=1}^{N_{L}} \sum_{j=1}^{N_{S}} w_{{\rm ds},ij} \times \gamma_{{\rm t},ij}\times \Sigma_{{\rm crit},ij}
    \over \sum_{i=1}^{N_{L}} \sum_{j=1}^{N_{S}}w_{{\rm ds},ij}},
\label{ds_equation}
\end{equation}

\noindent where $N_{L}$ is the number of lens galaxies and $N_{S}$ is the number of source galaxies. Each lens contributes a different effective weight in this sum. This topic is discussed further in Appendix \ref{app:weighting}.

We compute $\Delta\Sigma$ in 13 logarithmically spaced radial bins from $R_1=0.04$ \h Mpc to $R_2=15$ \h Mpc. The limit on our outer radial bin is set by the size of our bootstrap regions and is discussed in the following section.

\subsection{Bootstrap Errors and Variance in the Lensing Signal}\label{errors}

The covariance matrix and correlation matrix for our data vector of
$\ds$ values will be noted as ${\bf C}$ and ${\bf C}_\mathrm{corr}$
respectively. We estimate ${\bf C}$ from the data using stratified
bootstrap. Our bootstrap errors should account for the effects of correlated shape noise as well as for field-to-field variance in the lensing signal.

We divide CFHTLenS and CS82 into 74 roughly equal area
bootstrap regions (45 for CS82 and 29 for CFHTLenS). Each region is
$\sim$3 - 4 deg$^2$ which corresponds to regions with transverse
comoving dimensions of order 40-60 $h^{-1}$ Mpc at the redshift of the
CMASS sample. Our bootstrap regions are designed as a compromise
between two competing requirements. First, in order to ensure that the
bootstrap samples are independent, we require the size of the bootstrap
regions to be larger than the maximum scale used in the measurement
(15 $h^{-1}$ Mpc). Second, we need a large number of bootstrap regions
in order to reduce the noise in our evaluation of the covariance
matrix. However, this second requirement goes in the direction of
requiring many regions, which will thus necessarily have to be
smaller. Satisfying these two requirements determines the
maximum scale to which we compute our g-g lensing signal. Unless specified otherwise,
errors on our g-g lensing signals are derived using 10,000 resamplings of these bootstrap regions.

Although CFHTLenS and CS82 are fairly large surveys\footnote{For example the volume probed by CS82 over the range $0.43<z<0.7$
  (after subtracting masked out areas) corresponds to 0.0497 $h^{-3}$
  Gpc$^3$.}, we find that there is still a large field-to-field
variance in the amplitude of the CMASS lensing signal\footnote{\textcol{Inhomogeneity in the CMASS sample selection due to seeing and stellar density \citep[][]{Ross:2012,Ross:2016} may contribute to this variance and will be explored in future work.}}. To highlight
this fact, we compute the CMASS g-g lensing signal separately for W1
and W3 as well as for three independent Stripe 82 patches that roughly
match the areas of W1 and W3. Each of these patches contains about
4000 CMASS galaxies. Figure \ref{samplevariance} presents the CMASS
g-g lensing signal for each of these five sub-regions. As can be seen
from Figure \ref{samplevariance}, there is a significant amount of
variance between the g-g lensing signals of each of these five
independent patches. However, importantly, there is no obvious
systematic trend \emph{between} CFHTLenS and CS82 patches. This
test suggests that differences between the lensing signals from CFHTLenS
and CS82 can be attributed to field-to-field variance and not
systematic effects between the two surveys.

Our bootstrap errors should account for sample variance effects. However, because our analysis is based on a sub-region of the full BOSS footprint, possible large scale variations in the properties of the CMASS sample may be a concern. Figure \ref{samplevariance2} shows that the number density of CMASS within the CS82 footprint closely follows the number density of the full CMASS sample. We conclude from Figure \ref{samplevariance2} that substantial differences between the CS82 CMASS sample and the full DR12 CMASS sample are an unlikely possibility.

\begin{figure*}
\begin{center}
\includegraphics[width=17cm]{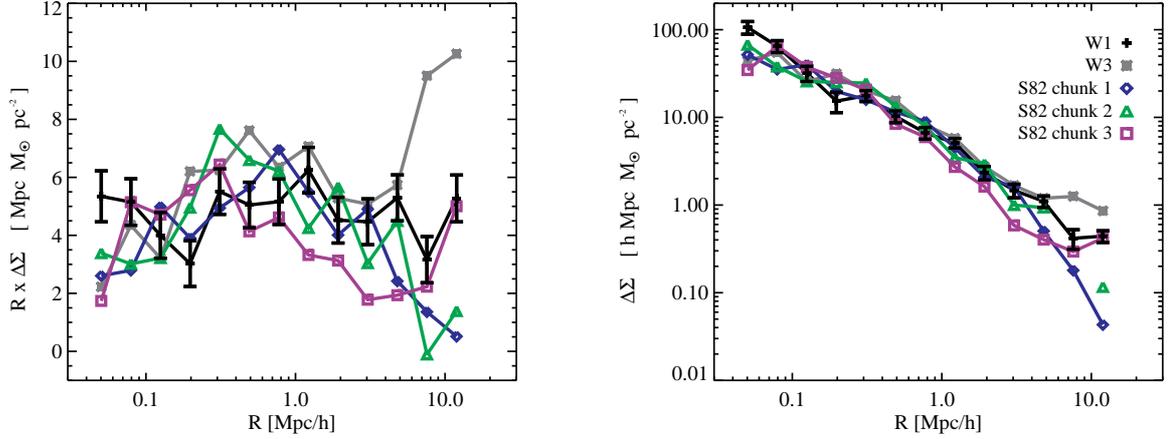}
\caption{CMASS g-g lensing signal calculated in five
  independent sub-regions of roughly equal area (2 regions from CFHTLenS
  and 3 from CS82). Each patch is of order $\sim$ 40-50
  deg$^2$. Because there are not enough bootstrap regions per
  sub-region to compute bootstrap errors, we simply show the typical
  shape noise errors for one of the patches (black data points) which underestimates the errors on large scales. As can
  be seen, there is a significant amount of field-to-field variance
  between the five independent patches. However, importantly, there is
  no obvious systematic trend \emph{between} CFHTLenS and CS82 patches.}
\label{samplevariance}
\end{center}
\end{figure*}

\begin{figure}
\begin{center}
\includegraphics[width=8cm]{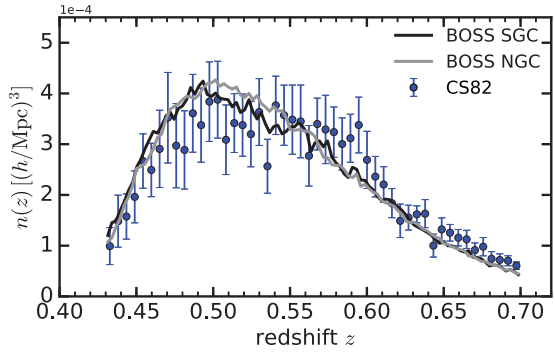}
\caption{Number density of CMASS galaxies within the CS82 footprint compared to the number density for the full DR12 sample. Black and grey solid lines show the $dn/dz$ for the North and South galactic cap respectively. Errors reflect the variance of $dn/dz$ between 24 independent BOSS patches that each have the same area as CS82.}
\label{samplevariance2}
\end{center}
\end{figure}

\subsection{Combined Signal from CS82 and CFHTLenS}\label{combinedsignal}

We first compute the CMASS g-g lensing signal separately for
CFHTLenS and CS82. The multiplicative shear calibration factor is
applied separately for each survey\footnote{The calibration factor is applied by dividing $\Delta\Sigma$ by $1+m$ where $m$ is the multiplicative shear calibration factor.}. Because the CS82 source
catalog is limited by \photozs and not shape
measurements, CS82 source galaxies have a higher mean signal-to-noise than CFHTLenS source galaxies. As a result, CS82 has a
smaller overall calibration factor compared to CFHTLenS. For CFHTLenS,
$1+m_{\rm cfhtls}\sim0.93$ which results in a 7.5\% increase in
$\Delta\Sigma$. For CS82, $1+m_{\rm cs82}\sim0.97$ which results in a
3\% increase in $\Delta\Sigma$.

Figure \ref{cs82cfhtls} displays the CMASS g-g lensing signal from
CFHTLenS and CS82. The signals agree well on small scales but there is
moderate amplitude difference at large scales which we attribute to
field-to-field variance as discussed in the previous section. Because
we do not have enough bootstrap regions to compute resampling errors
for CFHTLenS and CS82 separately, the errors displayed in Figure
\ref{cs82cfhtls} correspond to shape noise errors which will
underestimate the true variance on large scales. Also, from the
combined analysis, we expect the 5 outer points in these g-g lensing
signals to be moderately correlated (see Figure
\ref{covarmatrix}). Given this caveat, it is difficult to ascertain
the exact significance of the large-scale amplitude difference between
the two surveys. Instead, what we take away from
Figure \ref{cs82cfhtls} is that there is no evidence for a global
amplitude shift between the lensing signals from CFHTLenS and CS82.

\begin{figure}
\begin{center}
\includegraphics[width=8.5cm]{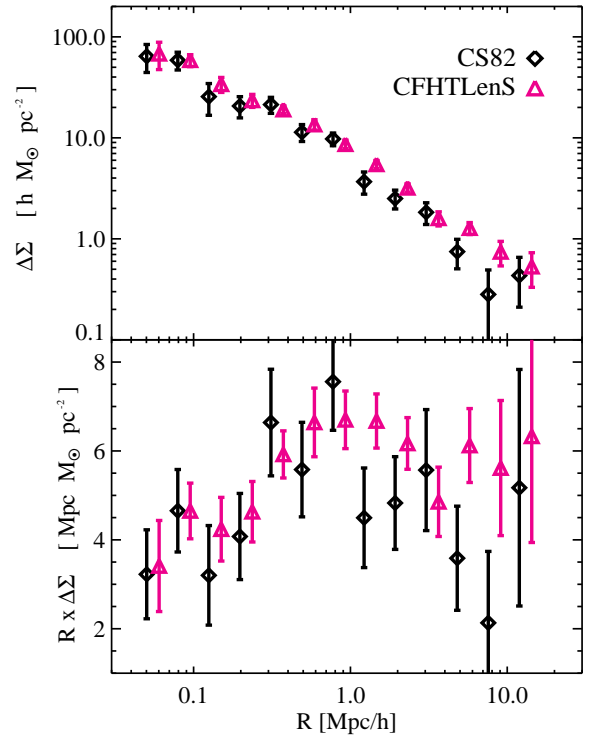}
\caption{Comparison between the CMASS g-g lensing signal from CS82
  (black diamonds) and from CFHTLenS (magenta triangles). The CFHTLenS data
  point are slightly offset for visual clarity. The signals agree well
  on small scales but there is moderate amplitude difference at larger
  scales which we attribute to field-to-field variance.}
\label{cs82cfhtls}
\end{center}
\end{figure}

Having convinced ourselves from Figure \ref{samplevariance} that there
are no obvious systematic trends between the two surveys, we now
proceed to combine the g-g lensing signal from CS82 and
CFHTLenS. There are 13775 CMASS galaxies from CS82 and 10507 from CFHTLenS
that are included in the weak lensing stack. Importantly, by combining
the two surveys, we gain a wider area with which to compute bootstrap
errors on the combined signal (74 bootstrap regions for the combined
sample).

Figure \ref{combinedsignalfigure} displays the combined g-g lensing
signal. The combination of CFHTLenS and CS82 yields a high $S/N$ measurement of the lensing signal for
CMASS. The $S/N$ of the signal is:

\begin{equation}
\frac{S}{N} = \left({\bmath x}^T {\bf C}^{-1} {\bmath x}\right)^{1/2},
\end{equation}

\noindent where ${\bmath x}$ is the vector of $\ds$ values in each
radial bin and ${\bf C}$ is the covariance matrix. There are 13 data
points in our stack and the relative error on each data point is
10-20\%. Our overall lensing signal is detected with a signal-to-noise
of $S/N=30$. For comparison purposes, the $S/N$ of the g-g
lensing signals used in the cosmological analysis of \citet{Mandelbaum:2013} had $S/N\sim25$\footnote{Over the radial range 0.1-70 Mpc/h, including small scale information that was not used in the cosmological analysis, the $S/N$ of the \citet{Mandelbaum:2013} measurement is $S/N=36$.}. 

\begin{figure*}
\begin{center}
\includegraphics[width=18cm]{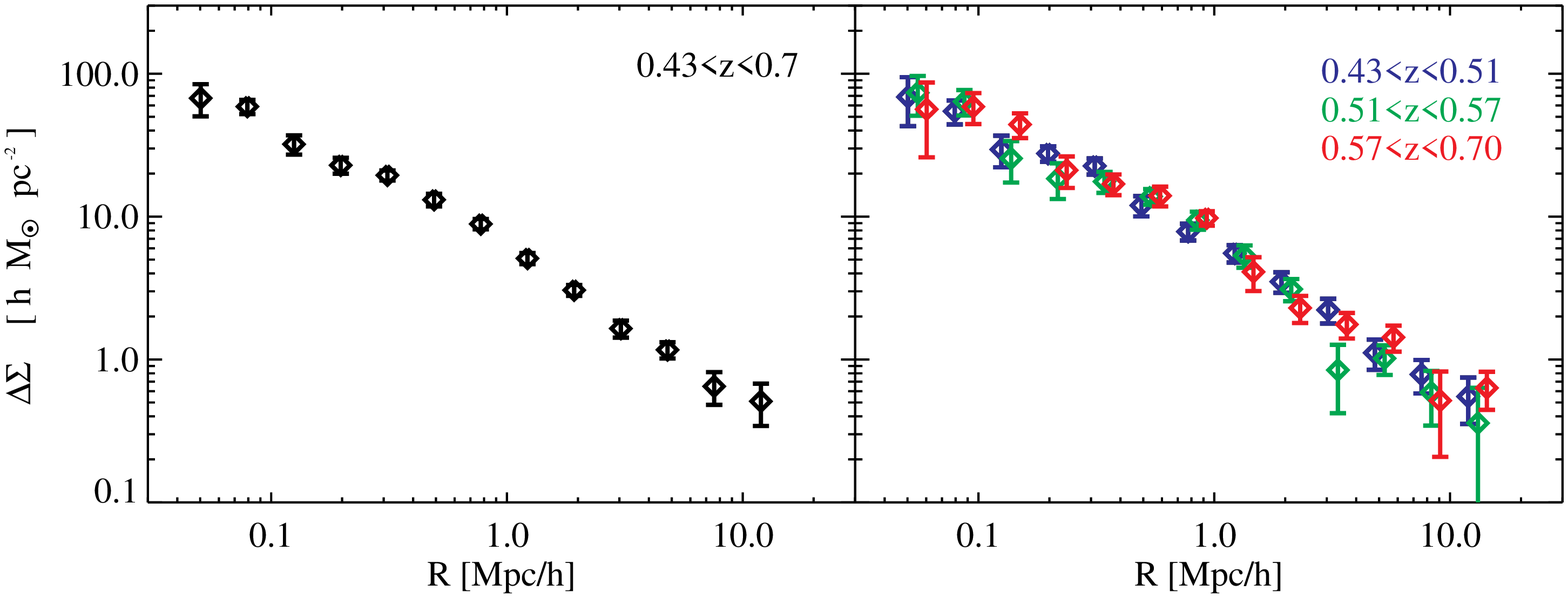}
\caption{Stacked weak lensing signal for CMASS using both CS82 and
  CFHTLenS. Left: lensing signal for CMASS in the redshift range
  $0.43<z<0.7$. Right: combined lensing signal for CMASS in three
  redshift bins. Data points in each redshift bin have been slightly
  offset for visual clarity. Errors are computed via bootstrap.}
\label{combinedsignalfigure}
\end{center}
\end{figure*}

We do not have enough bootstrap regions to constrain the full
covariance matrix, but we can constrain the dominant off-diagonal
terms. At small scales, we expect the data points to be uncorrelated \citep[\eg][]{Viola:2015}. On large scales, however, we expect non zero off-diagonal terms due to sample variance and correlated shape noise which arises when source galaxies appear in multiple radial bins \citep[][]{Jeong:2009aa,Mandelbaum:2013}. We compute the correlation matrix, ${\bf
  C}_\mathrm{corr}$, for our signal and then apply a boxcar smoothing
algorithm with a length of one bin in radius to this matrix, to reduce
the noise (see \citealt{Mandelbaum:2013} for a similar procedure). We
then truncate the correlation matrix to values greater
than 0.2 since we don't expect to constrain these terms\footnote{Our results are unchanged whether or not we apply the smoothing and truncation to ${\bf
  C}_\mathrm{corr}$.}. Because we reduce the noise in the correlation matrix directly, we do not attempt to apply any noise bias corrections when inverting ${\bf C}$ \citep[for example see][]{Hartlap:2007}. The result is displayed in Figure \ref{covarmatrix}. As expected, the outer data points are moderately correlated. The
dominant terms in the correlation matrix for the overall
signal are given in Table \ref{deltasigma_measure}.

\begin{figure}
\begin{center}
\includegraphics[width=8cm]{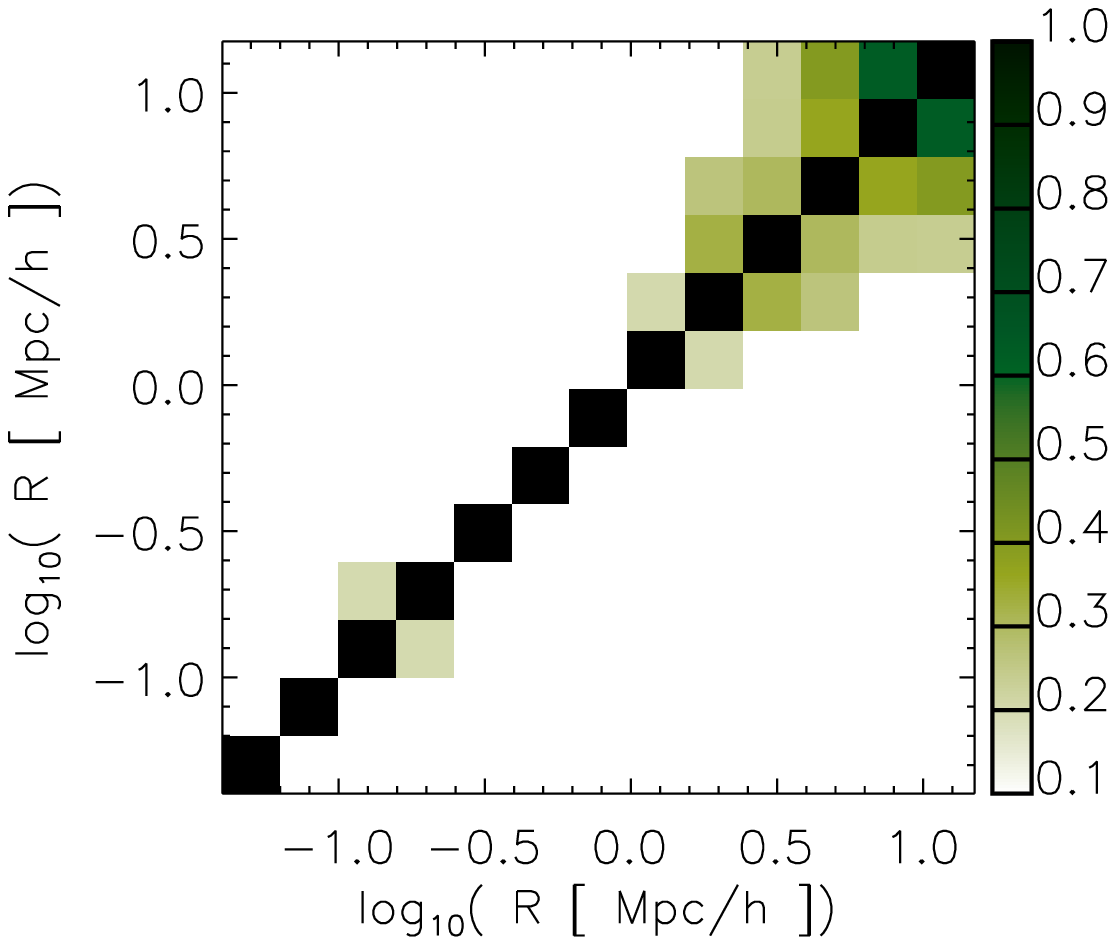}
\caption{Correlation matrix, $C_{\rm corr}$, computed via bootstrap,
  smoothed over one pixel scale, and truncated to values greater than
  0.2. The outer five data points in our g-g lensing are moderately correlated due to sample variance and correlated shape noise. The inner data points are uncorrelated and the errors on small scales are dominated by shape noise.}
\label{covarmatrix}
\end{center}
\end{figure}

We also compute the CMASS lensing signal in three redshift bins:
zbin1= $[0.43,0.51]$, zbin2=$[0.51,0.57]$, and zbin3=$[0.57,0.7]$. We
have checked that the multiplicative shear calibration factor, $m$,
does not vary strongly over this redshift baseline for either CFHTLenS
or CS82 and that the same calibration factor that we used for the
single wide redshift bin can be used for these more narrow redshift
bins. The g-g lensing signal for each of these redshifts bins is
presented in the right hand side of Figure \ref{combinedsignalfigure} and
the data points for our g-g lensing measurements are provided in Table
\ref{deltasigma_measure}. Interestingly, the amplitude of the g-g lensing
signal does not vary strongly with redshift -- we will return to this
point in Section \ref{discuss:evol}.

\begin{table*}
  \caption{Combined CS82+CFHTLenS g-g lensing measurements for CMASS. Errors are estimated via bootstrap (Section \ref{errors}). The
dominant terms in the correlation matrix for the overall
signal are: ${\bf C}_\mathrm{corr}[12,13]=0.64$, ${\bf
  C}_\mathrm{corr}[11,13]=0.42$, ${\bf C}_\mathrm{corr}[10,13]=0.25$,
${\bf C}_\mathrm{corr}[11,12]=0.38$, ${\bf
  C}_\mathrm{corr}[10,12]=0.25$, ${\bf C}_\mathrm{corr}[10,11]=0.30$,
${\bf C}_\mathrm{corr}[9,11]=0.26$, ${\bf
  C}_\mathrm{corr}[9,10]=0.32$. These values can be combined with the
errors given below to form the covariance
matrix for the signal measured over the full redshift range $0.43<z<0.7$. Note that these errors do not include a systematic uncertainty from photo-$z$s. In Section \ref{appendix_photoz} we estimate this systematic uncertainty to be of order 3\%. Our conservative estimate for the total fractional systematic error on $\Delta\Sigma$ is 5-10\%.}\label{deltasigma_measure}
\begin{tabular}{@{}cccccc}
\hline
Bin number &R [h$^{-1}$ Mpc] & $\Delta\Sigma$ [h M$_{\odot}$
pc$^{-2}$]&$\Delta\Sigma$  [h M$_{\odot}$ pc$^{-2}$] & $\Delta\Sigma$  [h M$_{\odot}$ pc$^{-2}$]&
$\Delta\Sigma$  [h M$_{\odot}$ pc$^{-2}$]\\
&&$0.43<z<0.7$&$0.43<z<0.51$& $0.51<z<0.57$& $0.57<z<0.7$\\
\hline
1&0.05&$67.16\pm16.77$&$67.95\pm25.50$&$73.43\pm23.01$&$57.03\pm29.74$\\
2&0.08&$58.98\pm6.73$&$54.83\pm10.58$&$63.67\pm13.04$&$59.08\pm14.40$\\
3&0.13&$32.12\pm4.95$&$29.61\pm7.33$&$25.31\pm8.22$&$44.26\pm8.71$\\
4&0.20&$22.95\pm2.90$&$27.69\pm3.45$&$18.57\pm5.14$&$21.14\pm5.25$\\
5&0.31&$19.50\pm1.58$&$22.71\pm2.92$&$17.59\pm2.98$&$16.92\pm2.80$\\
6&0.49&$13.13\pm1.33$&$12.02\pm1.95$&$13.84\pm1.73$&$13.98\pm2.22$\\
7&0.77&$8.88\pm0.74$&$7.87\pm1.06$&$9.46\pm1.36$&$9.75\pm1.14$\\
8&1.22&$5.11\pm0.45$&$5.56\pm0.76$&$5.34\pm0.96$&$4.12\pm1.09$\\
9&1.93&$3.06\pm0.27$&$3.50\pm0.58$&$3.11\pm0.55$&$2.31\pm0.49$\\
10&3.04&$1.65\pm0.22$&$2.23\pm0.43$&$0.85\pm0.42$&$1.76\pm0.35$\\
11&4.80&$1.17\pm0.15$&$1.12\pm0.26$&$1.03\pm0.24$&$1.43\pm0.30$\\
12&7.57&$0.65\pm0.17$&$0.79\pm0.21$&$0.60\pm0.25$&$0.52\pm0.31$\\
13&11.94&$0.51\pm0.17$&$0.55\pm0.20$&$0.37\pm0.28$&$0.64\pm0.19$\\
\hline
\end{tabular}
\end{table*}


\section{Results: Comparison with Predictions from Models Trained on Galaxy Clustering}\label{results}

We now compare our lensing signal with predictions from mock catalogs tailored to match the clustering of CMASS \citep[][]{Reid:2014,
  Saito:2016,Rodriguez-Torres:2015aa,Alam:2016}. These mock catalogs
were created by independent teams, using a range of methodologies,
cosmologies, {\it N}-body simulations (with varying resolutions), and
were all designed to reproduce the clustering of CMASS on
the scales relevant for this work ($r<30$ \h Mpc). Two mock employ SHAM whereas others employ an HOD based
method. The cosmology of these mocks ranges between a WMAP 5 cosmology \citep{Komatsu:2009aa} with $\Omega_m=0.27$ and a Planck-like cosmology \citep[][]{Planck-Collaboration:2015aa} with $\Omega_m=0.31$. Table \ref{sims} summarizes the
parameters of the various N-body simulations used to generate our
predictions. We do not use the Quick Particle Mesh
\citep[QPM,][]{White:2014aa} or PATCHY \citep[][]{Kitaura:2014aa}
mocks from the BOSS collaboration because these do not have the
necessary resolution to reproduce the galaxy-matter correlation
function on the scales of interest.  We begin with an overview of the
mocks used for our comparison.

\begin{table*}
  \caption{Simulation parameters for BOSS mock catalogs used in Figure \ref{bossmocks}.}\label{sims}
\begin{tabular}{@{}lcccccc}
\hline
Parameter & R14 & R14  & S16  & S16 updated &
RT16 & A16 \\
& MedRes & HiRes & MDR1 & MDPL2&
BigMDPL & MedRes \\
\hline
$L_{\rm box}$ ($h^{-1}$ Mpc) & 1380 & 677.7 & 1000 & 1000 & 2500 & 1380\\
$N_{\rm p}$ & 2048$^3$ & 2048$^3$ & 2048$^3$ & 3840$^3$ & 3840$^3$ & 2048$^3$\\
$\Omega_m$ & 0.292 & 0.30851 & 0.27 & 0.31 & 0.307 & 0.292\\
$\sigma_8$ & 0.82 & 0.8288 & 0.82 & 0.82  & 0.829 & 0.82\\
$z_{\rm box}$ & 0.550& 0.547&0.534& 0.457, 0.523,0.592 &0.505, 0.547,
0.623 & 0.550 \\
\hline
\end{tabular}
\end{table*}

\subsection{Overview of CMASS Mock Catalogs}


\citet[][]{Reid:2014} performed a joint analysis of the projected
and the anisotropic clustering (monopole and quadrupole) of CMASS on
scales from 0.8 to 32 $h^{-1}$ Mpc. Their analysis was performed by
populating an {\it N}-body simulation at $z=0.55$ with mock galaxies based on a
standard HOD type prescription. A single redshift-independent HOD
model was assumed with a number density of $\overline{n}\sim
4\times10^{-4} (h^{-1}\mathrm{Mpc})^{-3}$. Their mock
catalogs were randomly down-sampled along one of the axis of the
simulation to match the CMASS $dn/dz$. This procedure assumes that
CMASS galaxies at all redshifts are a random subsample drawn from a
single population. \citet[][]{Reid:2014} performed fits using two different simulations: a ``MedRes'' N-body simulation
($\Omega_m=0.292$, $\sigma_8=0.82$) and a ``HiRes'' simulation with a
Planck cosmology
\citep[$\Omega_m=0.30851$, $\sigma_8=0.8288$,][]{Planck-Collaboration:2014}. We
compare with the predictions from the best-fit models for both simulations. \citet[][]{Reid:2014} also perform several tests to verify the
robustness of their results to extensions of the standard HOD
model. In one such test, they consider a scenario in which 20 per cent
of centrals in massive haloes are not CMASS selected galaxies
(labelled ``cen/sat'' test in their paper). In this test, a central
galaxy is not required for a given halo to host a satellite galaxy. We
compare with both the fiducial model from \citet[][]{Reid:2014} as
well as with the ‘cen/sat’ model but find that both models generate similar predictions for the lensing signal.

\citet[][]{Saito:2016} present a joint modeling of both the projection
correlation function of CMASS ($w_{\rm p}^{\rm CMASS}$) and of the galaxy
stellar mass function (SMF) using SHAM. To perform SHAM,
\citet[][]{Saito:2016} use the galaxy SMF\footnote{Using SHAM in the mass range relevant for CMASS requires a measurement of the total (all galaxies, not just CMASS) galaxy SMF down to stellar masses of roughly $\log_{10}(M_*/M_{\odot})\sim10.8$.} computed by
\citet[][]{Leauthaud:2016} from the Stripe 82 Massive Galaxy
Catalog\footnote{Publicly available at \url{www.massivegalaxies.com}}
\citep[{\sc s82-mgc},][]{Bundy:2015}. They account
for the stellar mass incompleteness of CMASS by
down-sampling mock galaxies according to their assigned stellar mass
to match the redshift dependent CMASS SMFs. The \citet[][]{Saito:2016}
analysis used a single snapshot ($z = 0.534$) from the publicly
available ``MDR1'' MultiDark simulation
\citep[][]{Prada:2012aa,Riebe:2013aa} with a flat WMAP 5 $\Lambda$CDM
cosmology \citep{Komatsu:2009aa}. The \citet[][]{Saito:2016} mock
catalogs simultaneously reproduce $w_{\rm p}^{\rm CMASS}$, the galaxy SMF,
as well as the redshift dependent CMASS SMFs (and hence also reproduce the overall CMASS number density as a function of redshift).

In addition, we also compare with an updated version of
\citet[][]{Saito:2016} which uses the MDPL2 simulation from the
MultiDark suite. MDPL2 has the same box size ($L_{\rm box}=1~h^{-1}~{\rm
  Gpc}$) as MDR1 but has an improved resolution compared to MDR1
($N_{\rm par}=3840^3$). The $\Lambda$CDM cosmology in MDPL2 is
consistent with \citet[][]{Planck-Collaboration:2015aa}. MDPL2 keeps snapshots more frequently
than MDR1 and has more snapshots covering the CMASS redshift range. As
opposed to the MDR1 mock, here we use three different snapshots
($z=0.457$, $z=0.523$, $z=0.592$) and compute an updated version of the
\citet[][]{Saito:2016} mock catalogs by abundance matching each of
these snapshots.

\citet[][]{Rodriguez-Torres:2015aa} also use SHAM to build a CMASS
mock catalog that is designed to reproduce the monopole of the
redshift-space correlation function. Their mock is created from a
light-cone built from 20 outputs of the BigMDPL simulation
\citep[][]{Klypin:2016aa} and accounts for the geometry of the BOSS
survey as well as for veto masks. To perform SHAM,
\citet[][]{Rodriguez-Torres:2015aa} compute the galaxy SMF using
stellar masses from the Portsmouth DR12 catalog
\citep[][]{Maraston:2013}. In a similar fashion to
\citet[][]{Saito:2016}, the stellar mass completeness of CMASS is
modeled by down-sampling mock galaxies to reproduce the observed
number densities as a function of stellar mass. To compute the lensing
predictions from this mock, we use snapshots at three different
redshifts $z=0.5053$, $z=0.5470$ and $z=0.6226$.

\citet[][]{Alam:2016} build a CMASS mock catalog using a
standard four parameter HOD prescription. Their mock is based on the
same ``MedRes'' simulation employed by \citet{Reid:2014} but they use
a different procedure for populating this simulation with mock
galaxies. Whereas \citet{Reid:2014} place satellite galaxies on
randomly selected dark matter particles, \citet[][]{Alam:2016} place
satellite galaxies following a Navarro-Frenk-White profile
\citep[NFW;][]{Navarro:1997}. Whereas \citet{Reid:2014} uses halos
identified via a spherical-overdensity method, \citet[][]{Alam:2016}
uses halos identified using a friends-of-friends method with halo masses adjusted following \citet{More:2011a}. The HOD
parameters used by \citet[][]{Alam:2016} are tuned to match the projected
correlation function, $w_p^{\rm CMASS}$.

Among the various studies considered here, \citet[][]{Saito:2016} and
\citet[][]{Rodriguez-Torres:2015aa} are the only two that explicitly
model the stellar mass incompleteness of CMASS as a function of
redshift. The main differences between the two approaches are: the
size of the {\it N}-body simulation (representing a trade-off between
volume and resolution), the methodology for including scatter between
galaxy mass and halo mass in SHAM, and the origin of CMASS stellar
masses. In particular, the choice of a stellar mass estimator can lead
to important differences in the galaxy SMF (see Figure 15 in
\citealt[][]{Leauthaud:2016} for example). Both studies adopt $V_{\rm peak}$
(halo peak circular velocity) to perform SHAM. Both models account for
fiber-collision effects, either in the measurements themselves
\citep[][]{Saito:2016} or in the model
\citep[][]{Rodriguez-Torres:2015aa}. Importantly, the downsampling
procedure adopted in both studies assumes that CMASS galaxies are a
random sample of the overall population at fixed stellar
mass. However, \citet[][]{Leauthaud:2016} show that at fixed stellar
mass, CMASS is not a random sample of the overall population in terms
of galaxy color. In short: both methodologies account for mass incompleteness but not for color incompleteness. We will return to this point in Section \ref{sect:assemblybias}.

\subsection{Computation of Predicted Lensing Signal from Mocks}\label{computepredictions}

To compute the lensing signal predicted by CMASS mocks, we cross-correlate the positions of
mock galaxies with the positions of dark matter particles to form the
three dimensional galaxy-mass cross-correlation function, $\xi_{\rm
  gm}$. To compute $\Delta \Sigma$ from $\xi_{\rm gm}$ we follow the
equations outlined in section 4.2 of
\citet[][]{Leauthaud:2011}. Briefly, we begin by numerically integrating
$\xi_{\rm gm}$ over the line-of-sight to form the projected surface
mass density, $\Sigma$. In this step, it is important to perform the
integral out to a large radii or else $\Sigma$ will be underestimated. We find that integrating to 70-100 $h^{-1}$ Mpc is sufficient for
our purpose. Once we have computed $\Sigma$, we then compute $\Delta\Sigma$ via
two additional integrals -- details can be found in
\citet[][]{Leauthaud:2011}. We have verified that our code yields the
same result as an independent derivation using the {\sc halotools}
software package \citep[][]{Hearin:2016}. Finally, to account for the
contribution of the stellar mass of the galaxy to the lensing signal, we add a
point-source term to $\Delta\Sigma$ assuming a value of
$\log(M_*)=11.4$. This corresponds to the mean stellar mass of the
CMASS sample as computed from the {\sc s82-mgc}. In practice, this
point source term only has a minor contribution to $\Delta\Sigma$ at
$r<100$ \h kpc.

\subsection{Comparison between Predicted and Measured Lensing Signal}

Figure \ref{bossmocks} displays our main result which is the
comparison between the predictions from CMASS mocks and the measured lensing signal. All the predictions are
drawn from mocks which have a volume that is larger than the volume
corresponding to our lensing measurement. Hence, we neglect sampling
errors on the mock predictions which should be sub-dominant compared
to the errors on the measured lensing signal. The clustering
measurements used to construct these mocks were derived from a larger area than the lensing (thousands of square degrees compared to
a few hundred). Hence, any cross-covariance between the clustering and
lensing should be negligible.

The first point to take away from Figure \ref{bossmocks} is that all the mocks yield a surprisingly similar prediction for
$\Delta\Sigma$ with differences that are at most at the 20 per cent
level (with most models agreeing at the 15 per cent level). This is
quite remarkable given significant differences in the methodologies,
cosmologies, and {\it N}-body simulations used to construct
the mocks. In addition, each mock was tuned to match a different set
of observables -- some reproduce the projected correlation function
while others were tuned to fit the monopole or the quadrupole of the
three dimensional redshift-space correlation function. We conclude from Figure \ref{bossmocks} that, under standard assumptions about how
galaxies populate dark matter halos, the clustering of CMASS makes a
robust prediction for the amplitude of the lensing signal.

We now turn our attention to the comparison between the measured and
the predicted lensing signal. Figure \ref{bossmocks} shows that {\it
  all the mock catalogs} predict a lensing amplitude that is larger by
  $\sim$ 20-40\% than our measurement. For example, the $\chi^2$ between the measured lensing signal and the prediction from \citet[][]{Saito:2016} is $\chi^2/d.o.f=12.9$ with $d.o.f=13$. The $\chi^2$ for the updated \citet[][]{Saito:2016} MDPL2 mock is $\chi^2/d.o.f=14.1$. The $\chi^2$ difference with respect to other BOSS mocks have similar values.

\begin{figure*}
\begin{center}
\includegraphics[width=17.8cm]{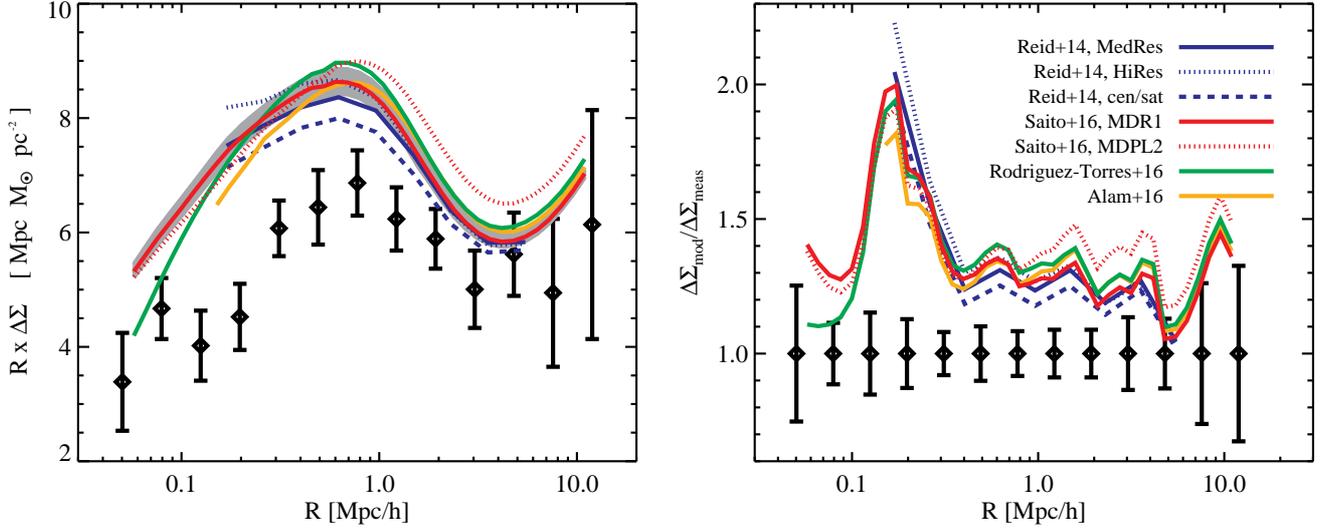}
\caption{Comparison of the g-g lensing signal with predictions from galaxy-halo models constrained by the clustering of CMASS. The grey shaded region represents models drawn from the 68\% confidence region for the \citet[][]{Saito:2016} MDR1 model. The ``spike" in the predictions in the right hand panel is simply cause by a downward fluctuation of the measured lensing signal at $r\sim 0.2$ \h Mpc as can be seen in the left panel. Regardless of the methodology (SHAM or HOD), the adopted cosmology, or the resolution of the $N$-body simulation, models constrained by the clustering of CMASS predict a lensing amplitude that is larger by $\sim$ 20-40\% than our measurement. This is not caused by different assumptions regarding $h$. The measurement and model predictions both assume a comoving length scale for $R$ and for $\Delta\Sigma$. Our code for computing $\Delta\Sigma$ yields the same result as an independent derivation by one of our co-authors. In Section \ref{app:compsdss} we show that CS82 lensing gives consistent results compared to SDSS. Finally, our code for computing model predictions yields the same result as the {\sc halotools}
software package \citep[][]{Hearin:2016}.}
\label{bossmocks}
\end{center}
\end{figure*}

Finally, we also investigate the redshift evolution of the CMASS g-g lensing signal. Figure \ref{shunmocksredshift} displays the lensing signal for CMASS in three redshift bins compared to predictions from \citet[][]{Saito:2016} and \citet[][]{Rodriguez-Torres:2015aa}. These results will be discussed in Section \ref{discuss:evol}.

\begin{figure*}
\begin{center}
\includegraphics[width=17.8cm]{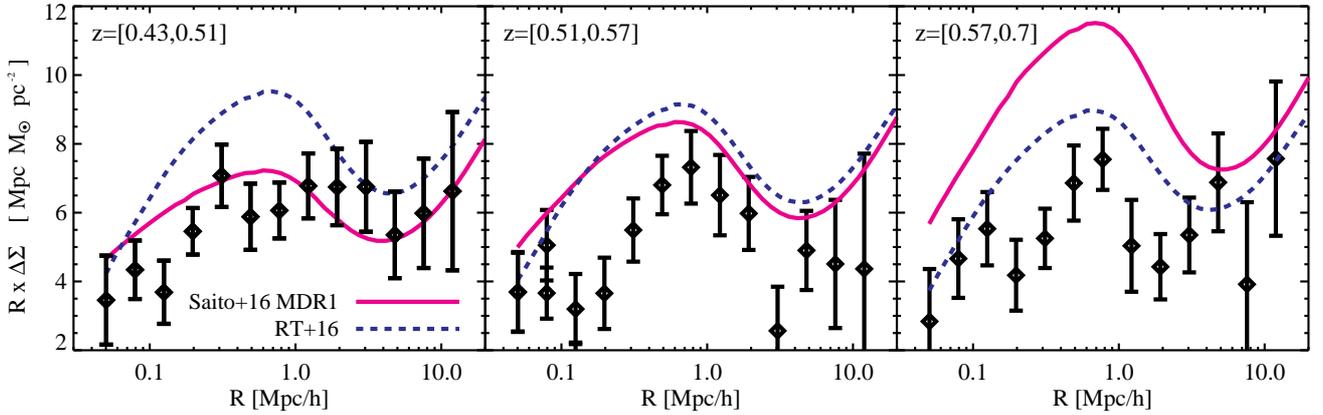}
\caption{Redshift evolution of the CMASS
  g-g lensing signal compared to predictions from
  \citet[][]{Saito:2016} and \citet[][]{Rodriguez-Torres:2015aa}. The
  \citet[][]{Saito:2016} model matches the lensing signal at low
  redshifts but then over-predicts the lensing signal at higher
  redshifts. The \citet[][]{Rodriguez-Torres:2015aa} model
  over-predicts the lensing signal by $\sim$ 20-40\% at all redshifts.}
\label{shunmocksredshift}
\end{center}
\end{figure*}


\section{Discussion}\label{discussion}

Our results demonstrate that standard models of the galaxy-halo connection tailored to reproduce the clustering of CMASS predict a g-g lensing signal that is 20-40\% higher than observed. We now discuss possible explanations for this mismatch. Because lensing measurements are non trivial, systematic effects are a concern. However, we argue below that the observed difference is too large to be explained by lensing related systematics effects alone. This leads us to consider other explanations including the impact of a low value of $\sigma_{8}$, sample selection effects and assembly bias, the impact of baryons on the matter distribution, massive neutrinos, and modified gravity effects. 

\subsection{Systematic Effects}\label{discuss:syst}

Could systematic effects explain the low amplitude of the lensing signal? Here we summarize and discuss the dominant effects which could impact our measurement. Further details on the various tests that we have performed can be found in the Appendices.

Our dominant source of systematic uncertainty is associated with the \photozs of source galaxies. If the \photozs of source galaxies are biased, this  may lead to a bias when evaluating the geometric factor $\Sigma_\mathrm{crit}$ (Equation \ref{eq:sigmacrit}). How much would the \photozs have to be wrong in order to explain Figure \ref{bossmocks}? It is difficult to give a succinct answer to this question because $\Sigma_\mathrm{crit}$ responds non linearly to $z_{S}$. However, to give an idea: when $z_L=0.55$ a 30\% effect on $\Delta\Sigma$  requires a source at $z_S=1$ to have a \photoz bias of $\Delta z=0.16$. Figure \ref{zphotzspec} which compares the \photozs of source galaxies from CS82 with known spectroscopic redshifts, shows no evidence for a bias this large. Furthermore, \citet[][]{Choi:2015} recently performed an analysis of the accuracy of the CFHTLenS \photozs and found at most at bias of $0.049$ in the photometric redshift bin spanning $0.57 < z_B < 0.7$ (in our case, most of our sources are removed from this range by our lens-source cuts). Finally, using a representative spectroscopic sample, we show in Appendix \ref{appendix_photoz} that the impact of \photoz errors on $\Delta\Sigma$ are at the 3\% level (this estimate includes the dilution of $\Delta\Sigma$ by source galaxies with $z_{S}^{\rm phot}>z_L$ but which are actually at redshifts below $z_L$). We conclude that \photoz bias alone is unlikely to explain Figure \ref{bossmocks}. 

It is common practice to apply a “boost correction factor” (see Appendix \ref{appendix_photoz}) to g-g lensing measurements to account for a dilution of the signal by physically associated sources. We have not applied this correction factor to our measurements for reasons that are outlined in Appendix \ref{appendix_boost}. In short, we argue that a variety of effects (masking, deblending, and failed photometry measurements in crowded regions) renders the computation of boost correction factors uncertain.  Instead, we adopt a more empirical approach and show that our lensing signal is robust to lens-source separation cuts (see Appendix \ref{appendix_photoz}). This test is based on the following argument: if the lensing signal is subject to a large dilution factor, then we expect the amplitude of the signal to increase for more conservative source selections. The fact that our lensing signal is invariant for a range of lens-source separation cuts suggest that dilution caused by physically associated galaxies is not a large concern. 

Another effect that we consider is that the weight function for CMASS is different between lensing and clustering measurements. Indeed, our predictions assume that lensing gives an equal weight to all halos, but there are a variety of reasons (outlined in Appendix \ref{app:weighting}) why this may not be true. However, in Appendix \ref{app:weighting} we show that the lensing signal is invariant even after removing the lensing specific weight function.

Finally, we consider the possibility of an unknown and unaccounted for bias in shear measurements from \emph{lens}fit. This question is of particular importance because 
other surveys which use \emph{lens}fit such as CFHTLensS and the KIlo-Degree Survey \citep[KIDS,][]{Kuijken:2015} report lower amplitudes for cosmic shear measurements \citep[\eg][]{Heymans:2013, Hildebrandt:2016} than predicted from Planck temperature fluctuations \citep[][]{Planck-Collaboration:2015aa}. Tests with image simulations suggest that the multiplicative bias for \emph{lens}fit is controlled to within a few percent \citep[][]{Fenech-Conti:2016}, but there is always the concern that shear calibration simulations may not be realistic enough. To address this concern, we measure the g-g lensing signal for a sample of massive low-redshift clusters from the redMaPPer cluster catalog \citep[v5.10, ][]{Rykoff:2014,Rozo:2014} and compare with a fully independent measurement using the SDSS catalog of \citet[][]{Reyes:2012aa}. The shear calibration method for \emph{lens}fit and for the \citet[][]{Reyes:2012aa} measurements are quite different: one uses simulations with galaxies described by simple sersic profiles, while the other is based on simulations with realistic galaxy morphologies drawn from Hubble Space Telescope (HST) imaging. Appendix \ref{app:compsdss} shows that the mean inverse-variance weighted offset between CS82 and SDSS is consistent with zero.  Furthermore, \citet[][]{Simet:2016} have shown that lensing measurements from \citet[][]{Reyes:2012aa} agree with yet another fully independent shear catalog referred to in their paper as the ``ESS'' catalogue \citep[\eg][]{Melchior:2014,Clampitt:2015}. The fact that three independent lensing measurements, with different shear calibration methods, yield the same results for $\Delta\Sigma$ suggest that a large bias in \emph{lens}fit shear measurements is an unlikely possibility.

In conclusion, while lensing is a difficult measurement to make, we conservatively estimate that the fractional systematic error on $\Delta\Sigma$ is less than 5-10\%. The differences reported in Figure \ref{bossmocks} are thus too large to be explained by systematic effects alone.
  
\subsection{Cosmology}

The predictions in Figure \ref{bossmocks} are generated from {\it N}-body simulations
with both WMAP and Planck-like cosmologies with $\Omega_{\rm m}$ values that span the range 0.27 to 0.31. However, as can be seen from Table \ref{sims}, the simulations used in our comparison only span the range $\sigma_8=0.82$ to $\sigma_8=0.829$. We now investigate how different $\sigma_8$ and $\Omega_{\rm m}$ would have to be in order to explain the the lensing signal.

So far, we have only considered model predictions derived directly from {\it N}-body simulations. There are two reasons for this. First, direct mock population provides a more robust theoretical prediction for our observables because, as opposed to analytic HOD type methods, there is no need to rely on analytical fitting functions for scale-dependent halo bias and halo exclusion. Second, as exemplified by \citet{Saito:2016} and \citet{Rodriguez-Torres:2015aa}, the idiosyncrasies of the CMASS sample (e.g., redshift-dependent selection effects which lead to a redshift dependent number density) can be directly folded into the modeling framework. The obvious downside of this approach, however, is that without resorting to sophisticated re-scaling \citep[\eg][]{Angulo:2015} or emulator type techniques \citep[\eg][]{Kwan:2015}, it is difficult to explore the cosmological dependencies of our observables. For this reason, we now adopt an analytic HOD model to investigate the cosmological implications of these lensing measurements.

Because of sample selection effects, we do not expect a single redshift independent HOD to capture the properties of CMASS (see Figure 10 in \citealt{Saito:2016}). However, our goal here is not to provide precision cosmological constraints, but simply to gain an intuition for the impact of cosmological parameters on $\Delta\Sigma$, and for this, a simple redshift independent HOD is sufficient. We use the analytical HOD modeling framework developed in \citet[][]{van-den-Bosch:2013aa}  to perform a joint fit to $\Delta\Sigma$ and $w_p^{\rm CMASS}$ (see \citealt{More:2013}, \citealt{Cacciato:2013}, and \citealt{More:2015} for an application of this method to SDSS data). This analytical framework accounts for the
radial dependence of halo bias, halo exclusion, residual redshift
space distortions in $w_p^{\rm CMASS}$, and the cosmological dependence of the
measurements \citep{More:2013aa}. For modeling the CMASS sample, we use a
simple 5 parameter description of the analytical halo occupation
distribution following \citet{Zheng:2007}, and a nuisance parameter
(see eq. 67 in \citealt{van-den-Bosch:2013aa}) that marginalizes over the
uncertainty in the model predictions near the one to two-halo
transition regime. We assume that the matter density within halos is
described by a NFW profile with the concentration-mass relation of \citet{Maccio:2008}. We also assume that the number of satellite
galaxies within halos of a given mass follows a Poisson distribution. Centrals are assumed to sit at the center of dark matter halos
while the number density distribution within the halo follows the dark
matter density.

Figure \ref{cosmologyplot} shows the contours on $\sigma_{\rm 8}$ and $\Omega_{\rm m}$ from our joint HOD fit compared to constraints from \citet{Planck-Collaboration:2015aa}\footnote{Specifically, we use Planck constraints that use both temperature and polarization data (Planck chain ``plikHM\_TTTEEE\_lowTEB'') as well as lensing of the CMB (Planck chain ``plikHM\_TTTEEE\_lowTEB\_lensing'').}. \textcol{Lensing plus clustering constrains the parameter combination $S_{\rm 8}=\sigma_8 \sqrt{\Omega_m/0.3}$. A 2-3$\sigma$ change in $S_{\rm 8}$ compared to Planck 2015 is required in order to match the lensing amplitude just via changes in cosmological parameters. When combined with external data such as BOSS BAO, and within the context of $\Lambda$CDM, WMAP 9 yields similar values for $S_{\rm 8}$ as Planck 2015. Hence, this may indicate a more general tension between g-g lensing and the CMB}. However, our measurements are dominated by highly non-linear scales, where other effects may also come into play (these will be discussed shortly), and so Figure \ref{cosmologyplot} should not be construed as direct evidence for a low value of $S_{\rm 8}$. Nonetheless, Figure \ref{cosmologyplot} does become more interesting when considered in the context of other independent constraints on the amplitude of low-redshift structure, both from lensing and from cluster abundances \citep[\eg][]{Heymans:2013,Planck-Collaboration:2015clusters, Hildebrandt:2016, Joudaki:2016,Giannantonio:2016}, that yield lower $\sigma_{\rm 8}$ values compared to Planck 2015 (but also see \citealt[][]{Jee:2013} and \citealt{The-Dark-Energy-Survey-Collaboration:2015} for cosmic shear results consistent with Planck 2015).

We comment on the fact at the cosmological constraints from \citet[][]{More:2015} show considerable overlap with the Planck constraints given the clustering and lensing signal of a subsample of CMASS. However, they have used a more flexible HOD, which includes a parametric form to model incompleteness, mis-centering of galaxies (or missing central galaxies with the CMASS selection criteria), and differences in galaxy and dark matter concentrations. Although such effects cannot be ruled out, this more flexible HOD leads to inflated inferred errors on cosmological constraints and thus may be hiding the discrepancy. We argue that characterizing and including sample selection effects into the modeling framework \citep[\eg][]{Saito:2016,Rodriguez-Torres:2015aa} is a more robust approach.

To summarize: lowering the value of $S_{\rm 8}$ by 2-3$\sigma$ compared to Planck 2015 reconciles the lensing with clustering. However, as argued in the following sections, at this level of precision, there are other effects that also come into play that need to be taken into consideration and disentangling these effects is a non trivial challenge.

\begin{figure}
\begin{center}
\includegraphics[width=8cm]{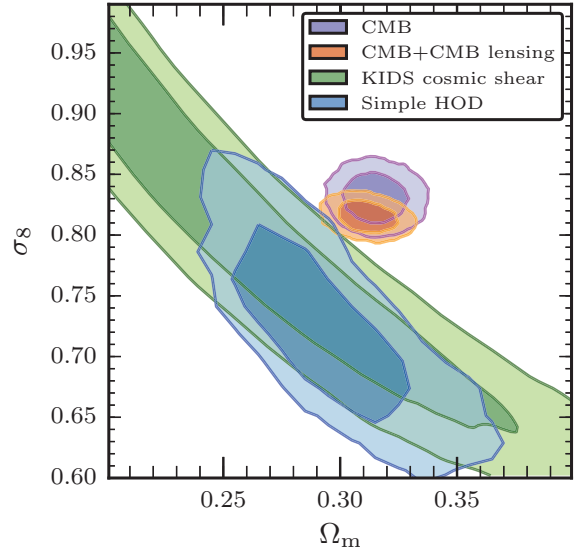}
\caption{Blue contours represent the result of a joint HOD fit to the two point correlation function and the g-g lensing signal for CMASS galaxies where $\Omega_{\rm m}$ and $\sigma_{\rm 8}$ are left as free parameters.  The lensing amplitude can be matched by lowering the value of $S_{\rm 8}$ which reduces the halo masses of galaxies at fixed number density. Green contours represent contours from the KIDS cosmic shear analysis of \citet[][]{Hildebrandt:2016}. Magenta contours show constraints from \citet[][]{Planck-Collaboration:2015aa} and orange contours include the CMB lensing effect. Lowering the value of $S_{\rm 8}$ compared to Planck 2015 reconciles the lensing with clustering.}
  \label{cosmologyplot}    
\end{center}
\end{figure}

  
\subsection{Sample Selection Effects}\label{discuss:evol}

On the relatively small radial scales considered here, the comparison between lensing and clustering is sensitive to the details of exactly how galaxies occupy dark matter halos. We now turn our attention to galaxy-formation related explanations for the low lensing amplitude.

The lack of redshift evolution of the lensing signal may contain important clues. The CMASS sample is not a single homogenous population and has properties that vary with redshift. According to the {\sc s82mgc}, the mean stellar mass of CMASS increases by a
factor of 1.8 over the range $0.43 < z < 0.7$. Based on the SHAM modeling of \citet[][]{Saito:2016}, this should lead to a factor of 3.5 increase in the predicted mean halo
mass\footnote{The redshift range of CMASS only corresponds to a time span of 2 Gyr and we do not expect much intrinsic evolution in the global connection
between galaxy mass and halo mass over such a short timeframe.} of CMASS from $z=0.43$ to $z=0.7$ (see Figure 12 in \citealt{Saito:2016}). This prediction stands in sharp contrast with the lack of redshift evolution in the CMASS lensing signal\footnote{The clustering of CMASS is also constant with redshift. See Figure A1 in \citet[][]{Reid:2014} and Figure 12 in \citet[][]{Saito:2016}.} displayed in Figure \ref{redshiftevol} and indicates that the models are an insufficient description of the data. One possible explanation for Figure \ref{redshiftevol} is that the mean stellar mass of CMASS evolves less strongly with redshift than predicted by the {\sc s82mgc}. For example, it is possible that the luminosity estimates from the {\sc s82mgc} have a redshift dependent bias because they do not fully capture light at the outskirts of galaxies. This type of bias could depend on galaxy type. New deep surveys such as HSC will yield better estimates for the total luminosities of massive galaxies and will shed light on this question (Huang et al in prep).

Another important point is that although some of the mocks discussed so far account for the stellar-mass completeness of the sample, none account for color completeness \emph{in addition} to mass completeness. The color-cuts that define CMASS exclude galaxies at low redshift with recent star formation. At higher redshifts ($z>0.6$), the sample is mainly flux-limited and includes a larger range of galaxy colors at fixed magnitude (see Figure 5 in \citealt{Leauthaud:2016}). A range of studies suggest that at fixed stellar mass, galaxies with different levels of star formation live in halos of different mass. At low redshift, studies find that at fixed stellar mass, blue central galaxies live in
lower mass halos \citep[\eg][and references therein]{Mandelbaum:2016}. At higher redshift, there are suggestions that this
trend may reverse \citep[][]{Tinker:2013}. A possible explanation of Figure \ref{redshiftevol} is that the inclusion of
more blue galaxies in the CMASS sample at higher redshifts leads to a
coincidental compensation that keeps the amplitude of the lensing fixed. However, although this may explain the lack of evolution in the lensing -- this does not immediately explain why the predicted lensing signal is lower than observed unless CMASS galaxies occupy halos in a way that leads to an unusual\footnote{Unusual here means unlike the range of models considered in Figure \ref{bossmocks}.} relation between the mass of their dark matter halos and their large scale clustering properties. For example, assembly bias may be at play and is discussed in the next section.

\begin{figure}
\begin{center}
\includegraphics[width=9cm]{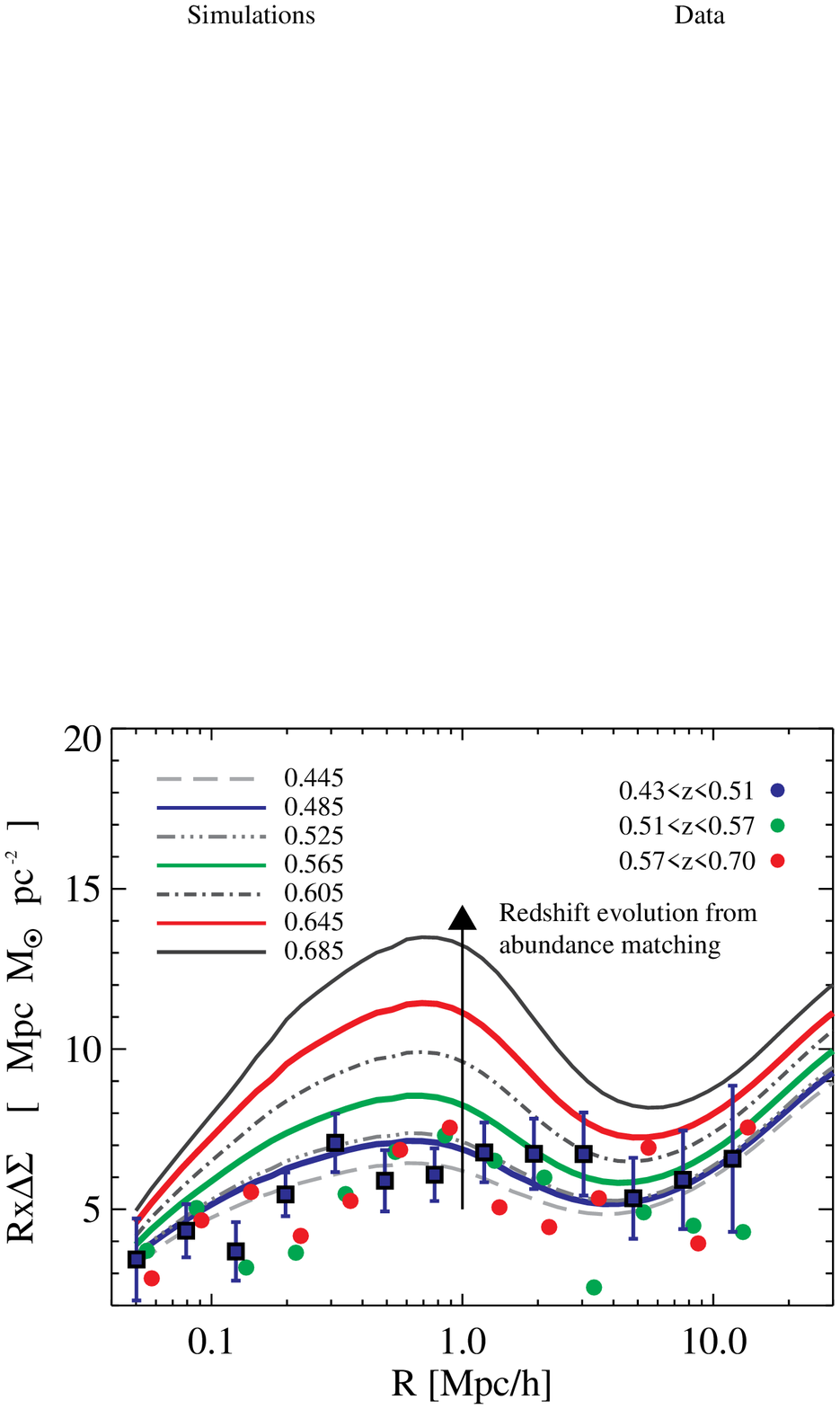}
\caption{Predicted redshift evolution of the lensing signal from the abundance matching model of \citet[][]{Saito:2016} compared to the
  measured lensing amplitude in three redshift bins. According to the
{\sc s82mgc}, the mean stellar mass of the CMASS sample increases by a factor of 1.8 over the range $0.43<z<0.7$. As a consequence, SHAM predicts that the mean halo mass should increase by a factor of 3.5 from low to high redshift.}
\label{redshiftevol}
\end{center}
\end{figure}

\subsection{Assembly Bias Effects}\label{sect:assemblybias}

The model predictions shown in Figure \ref{bossmocks} use standard galaxy-halo modeling based on either HOD or SHAM type methodologies. The fact that the amplitude of the lensing does not match the predictions from these models may reflect an inherent failure of such models. In particular, as highlighted by \citet{Zentner:2014}, one aspect that has recently come to the forefront is that these models\footnote{Standard HOD models have no assembly bias, whereas SHAM models based on $V_{\rm peak}$ do have some levels of assembly bias \citep[][]{Zentner:2014}.} neglect assembly bias: the fact that in addition to halo mass, the strength of halo clustering depends on other properties such as halo age, spin, and concentration \citep[][]{Gao:2005,Wechsler:2006,Gao:2007, Zentner:2007, Dalal:2008,Lacerna:2011}. Whereas assembly bias is manifest in dark matter simulations, we do not know if it is also manifest in the clustering of galaxies. Recent observational evidence suggests the possibility of assembly bias in galaxy and cluster samples \citep[][]{Lehmann:2015, More:2016, Miyatake:2016, Zentner:2016}, but these detections are not without challenges \citep[][]{Paranjape:2015,Lin:2016}.

If galaxy formation processes are sensitive to halo parameters besides halo mass, for example, if the ages of galaxies correlate with the ages of their dark matter halos, then assembly bias effects will be more pronounced for color selected samples such as CMASS. The clustering of CMASS tightly constrains the large-scales bias of the sample. However, the lensing signal that we measure is limited to $r<10$ \h Mpc and is primarily sensitive to the one-halo term and the mean halo mass of the sample. Hence, the difference that we observe may suggest a tension between the halo mass and the large-scale bias of this sample -- the smoking gun for assembly bias. This interpretation would mean that CMASS host halos are not a representative sample of all dark matter halos at the same mass, and since the bias of halos depends on other properties apart from their mass, they thus show a different clustering amplitude than such a representative sample.

In \citet{Saito:2016} we present the first analysis of the effects of assembly bias on the clustering properties of CMASS. However, our analysis assumed a simplified model for the color completeness of CMASS. To build on \citet{Saito:2016}, the next step would be to characterize the color-completeness of CMASS and to explore the impact of assembly bias using, for example, conditional subhalo abundance matching techniques \citep[\eg][]{Hearin:2014a}. \textcol{This type of in-depth analysis is beyond the scope of this paper. Instead, we present a simple first-order computation to determine if assembly bias is a plausible explanation for the observed offset. We fit a simple four parameter HOD to $w_{p}^{\rm CMASS}$ (details are given in Appendix \ref{app:hodfit}) and show the results in Figure \ref{abtest}. The predicted lensing signal can be decomposed into three components: the one-halo central term ($\Delta\Sigma_{\rm 1hc}$), the one halo satellite term ($\Delta\Sigma_{\rm 1hs}$), and the two-halo term ($\Delta\Sigma_{\rm 2h}$). Figure \ref{abtest} shows that the amplitude of the lensing signal is well matched if the one-halo central term is decreased by 25 per cent while keeping the two halo term fixed\footnote{This exercise is simplistic because it does not necessarily preserve the CMASS clustering or abundance.}. In this regime, $M_{\rm halo} \propto (\Delta\Sigma_{1hc})^{3/2}$ so this corresponds to a $\sim$35 per cent decrease in halo mass. The halo masses of CMASS galaxies are firmly above collapse mass at $z=0.55$ \citep[][]{Saito:2016} where the effects of assembly bias are complex\footnote{The magnitude and sign of assembly bias effects above collapse mass depends sensitively on the definition of halo age \citep[\eg][]{Li:2008}.} and not yet necessarily well characterized. With this caveat in mind, assembly bias can plausibly explain a $\sim$35 per cent decrease in halo mass at fixed bias (see Figure 4 in \citealt[][]{Li:2008} for example). Lensing measurements on larger radial scales will be extremely valuable for testing this hypothesis.}

If assembly bias is at play, this could have implications for growth of structure constraints from redshift-space distortions \citep[RSD,][and references therein]{Alam:2016a}. Unlike BAO measurements, RSD methods push into the semi-non linear regime and need to be validated against using mock catalogs. Current tests suggest that RSD methods are robust to the details of galaxy formation (see Section 7.2 in \citealt{Alam:2016a}), but the full range of galaxy formation models has yet to be tested, and hence the theoretical systematic associated with the complexities of galaxy bias is unknown. Assembly bias and the details of the galaxy-halo connection may become an important systematic effect for RSD constraints from upcoming surveys such as DESI \citep{Levi:2013}. Lensing measurements such as presented in this paper will play an important role in characterizing these effects.

 \begin{figure*}
\begin{center}
\includegraphics[width=16cm]{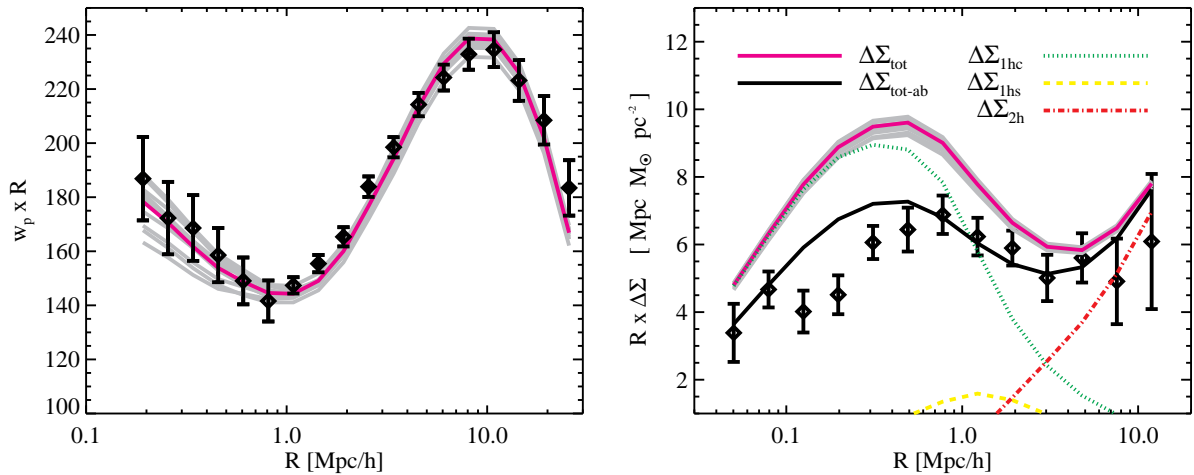}
\caption{Right panel: a simple four parameter HOD fit to $w_{p}^{\rm CMASS}$ at fixed cosmology. Grey lines represent models drawn from the the 68\% confidence region. Right panel: predicted lensing signal (solid magenta line). Grey lines represent models drawn from the 68\% confidence region of the best-fit to $w_{p}^{\rm CMASS}$. The lensing signal can be decomposed into a one-halo central term (green dotted line), a one-halo satellite term (dashed yellow line), and a two-halo term (red dash dot line). The satellite fraction for CMASS is only of order $\sim$10 percent and the one-halo satellite term is therefore sub-dominant on all scales. The black solid line is the total lensing signal obtained by lowering $\Delta\Sigma_{\rm 1hc}$ by 25 per cent, which roughly corresponds to lowering the halo mass by 35 per cent while keeping the bias fixed.}
\label{abtest}
\end{center}
\end{figure*}

\subsection{Baryon Effects}
\label{sec:baryons}

The BOSS CMASS mock catalogs used for computing the model predictions are based on gravity-only $N$-body simulations, which do not account for possible effects of baryon physics processes on the matter distribution. However, baryon physics processes
can affect the matter profiles of halos and also influence the properties of subhalos \citep[\eg][]{van-Daalen:2014,Velliscig:2014,Chaves-Montero:2016}. 

We use the Illustris simulations \citep[][]{Vogelsberger:2014a, Vogelsberger:2014b, Genel:2014, Sijacki:2015, Nelson:2015} to estimate the impact of baryonic effects for CMASS-like samples. We compare results from snapshots at redshift $z=0.5$ of the full-physics Illustris-1 simulation and of the corresponding gravity-only Illustris-1-Dark simulation with matched initial conditions. The Illustris simulation corresponds to a comoving volume of (75 $h^{-1}$ Mpc)$^{3}$ which means that there will be considerable sample variance uncertainties associated with galaxy selections at these number densities. Our goal here, however, is not to compare directly with the BOSS measurements, but simply to estimate relative differences between the full-physics and gravity-only runs.

We rank order subhalos in both simulations according to their maximum circular velocity, $V_{\mathrm{max}}$, and apply a sharp lower limit on $V_{\mathrm{max}}$ to select samples with number densities of $\overline{n} = 4 \times 10^{-4} (h^{-1}\mathrm{Mpc})^{-3}$. The resulting lower limit is $V_{\mathrm{max}} = 351 \,\kms$ for the gravity-only run, and $V_{\mathrm{max}} = 367 \,\kms$ for the full-physics run. This selection results in 170 galaxies\footnote{With only 170 subhalos with $V_{\mathrm{max}} = 351 \,\kms$,  the Illustris simulation is not large enough to compute the clustering signal for galaxies at these low number densities. Our tests are therefore based on a simple number density selection without also matching the clustering.}. In addition to this sample which includes all sub-halos, we also perform a number density selection which includes only \emph{matched parent} halos. $\Delta\Sigma$ is computed for all samples using each of the three principal box axes as a viewing direction using Fast Fourier Transform methods \citep[][]{Hilbert:2011,Hilbert:2016}. Finally, $\Delta\Sigma$ is also computed from the gravity-only run with an added contribution from the stellar component computed from the full-physics run. The resulting weak lensing profiles are shown in Figure \ref{hydro_ds2}.

The upper panel in Figure \ref{hydro_ds2} show the impact of baryons on $\Delta\Sigma$ for matched parent halos with $\overline{n} = 4 \times 10^{-4} (h^{-1}\mathrm{Mpc})^{-3}$. For small separations ($R<0.1~h^{-1}\,\Mpc$), $\Delta\Sigma$ is larger in the full-physics simulation than in the gravity-only simulation. This is mainly due to the contribution from stars, which are missing in the gravity-only run. On intermediate scales, $\Delta\Sigma$ is larger in the gravity-only run than in the full-physics run by up to 20\%. This is due to feedback processes in the full-physics simulation that drive matter out of the inner parts of halos. These feedback processes also lower the baryon-fraction in the halos and decrease the matter power spectra on these scales \citep[][]{Vogelsberger:2014a}. However, $\Delta\Sigma$ converges on larger scales ($R > 4~h^{-1}\,\Mpc$), indicating that the impact of baryons for central halos is primarily limited to the one-halo regime.

The lower panels in Figure \ref{hydro_ds2} show $\Delta\Sigma$ for a fixed number density selection including all subhalos. The main difference with respect to the matched-parent sample is now that on large scales $\Delta\Sigma$ is higher in the full-physics run than in the gravity-only run. This is because the full-physics run has a larger satellite fraction ($f_{\rm sat}=$22\% compared to 11\% in the gravity-only run), and these satellites live is massive host halos. The larger fraction of satellites (and subhalos with large $V_{\mathrm{max}}$) is probably because satellites are more resistant to tidal stripping and are able to survive longer in the full-physics run\footnote{\textcol{Determining the subhalo fraction is a difficult task because simulations which lack resolution may result in artificial subhalo disruption. We raise this as a caveat to the numbers presented here, but have not explored these aspects further.}}. This factor of two difference in the satellite fraction between the full-physics run and the gravity-only run is particularly interesting because $w_p^{\rm CMASS}$ is very sensitive to this quantity. For example, the error on the CMASS satellite fraction from \citet[][]{Reid:2014} is less than one percent\footnote{\textcol{Our Illustris tests are performed using a simple fixed number density cut and are hence only a very loose approximation of the CMASS sample. Without a more careful attempt to match the stellar mass distribution of CMASS, the \citet[][]{Reid:2014} constraint on  $f_{\rm sat}$ ($10.16\pm0.69$\%) should not be directly compared with the values quoted for our Illustris sample.}}! If these constraints are robust, they could be very informative for feedback models. However, given the tensions with respect to the lensing, it is not clear if these HOD constraints on the CMASS satellite fraction are indeed robust. What is clear, however, is that a factor of two difference in the satellite fraction will have a large impact on CMASS abundance matching models which are currently based on gravity-only $N$-body simulations. \textcol{Analytic HOD models may be able to marginalize over baryonic effects by allowing the concentration of the satellite distribution and the concentration of the parent dark matter halo to vary as free parameters \citep[\eg][]{van-den-Bosch:2013aa,Reddick:2014}. But current implementations of SHAM, and HOD models based on direct $N$-body mock population, do not have this flexibility}. 

The tests presented here suggest that baryonic effects can induce a 10-30\% difference in $\Delta\Sigma$\footnote{As we were finishing this paper, new simulations from the Illustris group with an improved AGN feedback model suggest a smaller impact of baryonic effects on halo masses \citep{Weinberger:2016}.} (with a characteristic scale-dependence) and a factor of two difference in the satellite fraction for galaxy samples with $\overline{n}=4\times10^{-4}$ $(h^{-1}$ Mpc$)^{-3}$. This level of difference is no longer negligible given the statistical errors on our measurements. Without further analysis, it is difficult to say exactly how these effects would play out in an HOD or SHAM analysis of the clustering of BOSS galaxies, and whether or not the differences go in the same direction as our lensing measurements. However, it is clear that these effects warrant further investigation.

\begin{figure*}
\begin{center}
\includegraphics[width=16cm]{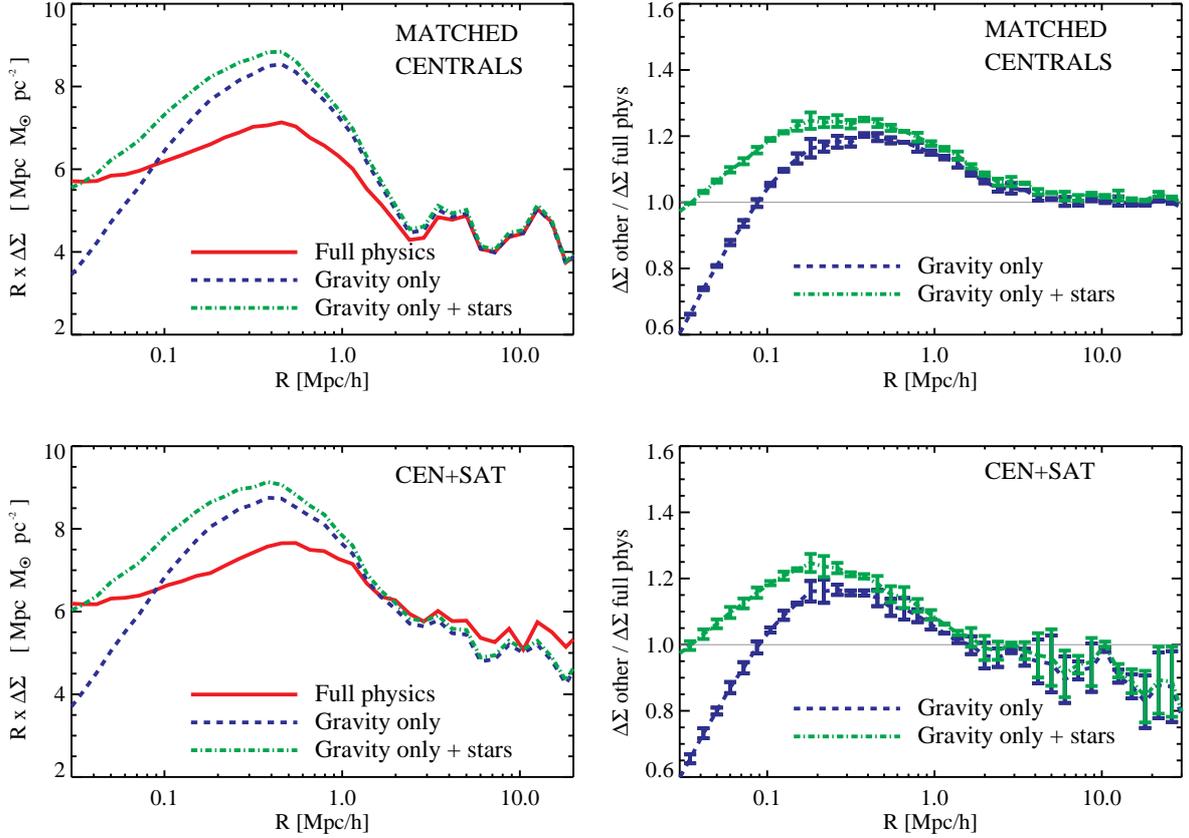}
\caption{Left panels: comparison of $\Delta\Sigma$ for massive galaxies with CMASS-like number densities for the gravity-only Illustris-1-Dark run (blue dashed line) and the full-physics Illustris-1 run (solid red line). The green dash-dot line shows $\Delta\Sigma$ for the gravity-only run plus the stellar component from the full-physics run. Right panels: Ratio of $\Delta\Sigma$ from the gravity-only run and from the full-physics run (blue line). The green dash-dot line shows the result including the contributions from stars. The error bars indicate the spread between the ratios obtained from using the three different principal simulation box axes as viewing direction. Upper panels: fixed number density selection for matched centrals. Lower panels: fixed number density selection including both centrals and satellites. }
\label{hydro_ds2}
\end{center}
\end{figure*}
 
\subsection{\textcol{Effect of Massive Neutrinos}}\label{neutrinos}
 
The total sum of neutrino masses is tightly constrained by cosmological observations 
to $\sum m_{\nu}<0.1$-$0.5$eV \citep[e.g.][]{Alam:2016a,Beutler:2014,Palanque-Delabrouille:2015,Saito:2011,Zhao:2013}. Finite-mass neutrinos have large velocity dispersion and suppress the growth 
of large-scale structure below the neutrino free-streaming scale 
\citep[e.g.,][]{Lesgourgues:2006,Saito:2008,Saito:2009} and could therefore impact the amplitude of the g-g lensing signal. Here we make a simple attempt to quantify the impact of neutrino masses on the g-g lensing signal (see also \citealt{Mandelbaum:2013} and \citealt{More:2013}). We run three $N$-body simulations using the particle-based method of \cite{Villaescusa-Navarro:2014} with initial conditions generated following \cite{Zennaro:2016}. The three simulations share the same initial seeds 
and have the same value for the total matter density ($\Omega_{\rm m}=\Omega_{\rm CDM}+\Omega_{\rm b}+\Omega_{\nu}=0.3175$) but have different neutrino masses (0eV, 0.15eV and 0.3eV). Our $N$-body simulations are created using {\sc Gadget-3} \citep{Springel:2005} with parameters $L_{\rm box}=300\,{\rm Mpc}/h$, $N_{\rm CDM}=512^{3}$, and $N_{\nu}=512^{3}$ (for the non zero neutrino mass simulations). Subhalos are identified using the SUBFIND algorithm \citep{Springel:2001,Dolag:2009} and a CMASS-like sample is selected via a simple constant number density cut with 
$\overline{n}=3\times 10^{-4}\,(h/{\rm Mpc})^{3}$ after rank-ordering subhalos by $V_{\rm max}$.

Figure \ref{nplot} shows the impact of massive neutrinos on $\Delta\Sigma$. A larger neutrino mass results in a lower amplitude for $\Delta\Sigma$, but only affects the signal at the $\sim 10\%$ level even with 0.3eV. This suppression is expected because massive neutrinos alter the halo mass function and globally reduce halo masses \citep[][]{Castorina:2014,Castorina:2015,Ichiki:2012}. Indeed, the mean halo masses for the three samples are $\log_{10}\overline{M}_{\rm vir}=13.60$, 13.55, and 13.51 for the 0eV, 0.15eV, and 0.3eV simulations respectively. Finally, we also find that the satellite fractions and the galaxy-galaxy correlation function are very similar among the three simulations. This is also expected because neutrinos only have a small impact on physics in the 1-halo regime and the difference in the 1-halo regime is mainly driven by differences in $\sigma_8$ \citep{Fontanot:2015}. 

We conclude from Figure \ref{nplot} that the effect of massive neutrinos goes in the right direction to explain the low amplitude of our lensing signal. However, the impact of massive neutrinos on $\Delta\Sigma$ is at the $\sim$10\% level at most, and so massive neutrinos alone are unlikely to be the full story.

\begin{figure}
\begin{center}
\includegraphics[width=8.5cm]{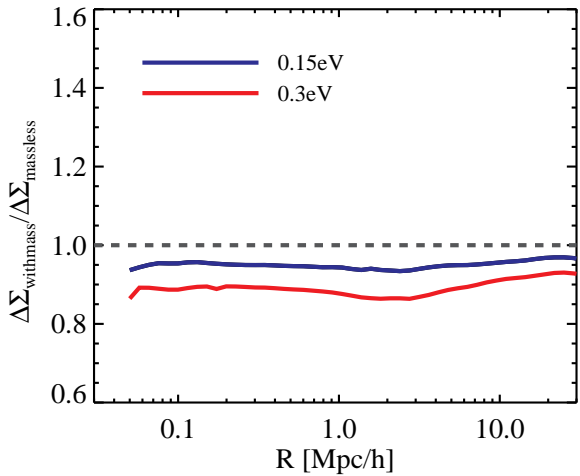}
\caption{Ratio of $\Delta\Sigma$ from simulations with massive neutrinos to $\Delta\Sigma$ from simulations with massless neutrinos. The blue corresponds to $\sum m_{\nu}=0.15$eV and the red line corresponds to $\sum m_{\nu}=0.3$eV. The impact of massive neutrinos leads a fairly scale independent decrease of $\Delta\Sigma$ over the scales of interest that could partially alleviate the tension reported in this paper.}
\label{nplot}
\end{center}
\end{figure}

\subsection{Modified Gravity Effects}\label{mgeffects}

Observations of redshift-space distortions provide an exciting opportunity to constrain models of modified gravity \citep[][]{Zhang:2007, Reyes:2010, Samushia:2013}. In particular, one promising method is to examine the velocity structure around massive clusters with halo masses determined via weak lensing \citep[][]{Schmidt:2010,Lombriser:2012,Lam:2012,Zu:2014}. If general relativity (GR) is valid, then the phase-space around clusters is uniquely determined by the mass of the clusters that source these velocities (but see \citealt{Hearin:2015} for caveats due to assembly bias effects). Although the CMASS sample is more complex than a simple cluster selection (which means that this test can only be carried out in tandem with the modeling of the CMASS-halo connection), differences between the lensing and predictions from models trained on the two dimensional redshift-space correlation function \citep[][]{Reid:2014,Rodriguez-Torres:2015aa} could be a signature of modified gravity. However, it is not immediately clear if deviations from GR would result in an {\em increase} or a {\em decrease} of the lensing amplitude.

To investigate this question, we use a suite of four $z = 0.57$ CMASS mock catalogs from \citet{Barreira:2016a}. One of these mocks is a $\Lambda$CDM control sample. The three other mocks are built from simulations (with $L_{\rm box}$=600 \h Mpc and $N_{\rm p}=1024^3$) of structure growth for the normal branch of the Dvali-Gabadadze-Porrati gravity model \citep[DGP,][]{Dvali:2000} which were created using the {\tt ECOSMOG} N-body code \citep{Li:2012, Li:2013}. The three DGP gravity mocks simulate strong, medium and weak departures from GR and will be referred to respectively as ``DGPs'', ``DGPm'', and ``DGPw''. The expansion rate in these simulations matches the $\lcdm$ control simulation which means that any differences compared to $\lcdm$ are induced by modifications to the gravitational force law \citep{Schmidt:2009}. CMASS mocks were created using an HOD model with parameters tuned to roughly match the CMASS number density and the large-scale amplitude of the CMASS power spectrum monopole. While these mocks were not designed to reproduce the clustering of CMASS as accurately as those used in Figure \ref{bossmocks}, they are still useful to understand the relative effects on $\Delta\Sigma$ for DGP-like models. Figure \ref{mgplot} shows the lensing signals of the DGP and $\lcdm$ CMASS mock samples. We find that DGP gravity leaves a scale-dependent signature in $\Delta\Sigma$ with a transition region located at $r\sim1$ \h Mpc. There are at least two relevant effects responsible for the difference between the DGP results and $\lcdm$ which are now discussed.

The first effect is due to the existence of a positive additional ``fifth" force in the DGP simulation which is common feature in many modified gravity models \citep[\eg][]{Joyce:2016}. At fixed halo mass, the fifth force favors the pileup and clustering of matter close to the accretion region of dark matter halos ($r> 1$ \h Mpc) which leads to a boost in the amplitude of the lensing signal. On smaller scales ($r< 1$ \h Mpc), the effects of the fifth force on matter clustering tend to become less pronounced because of the efficient Vainshtein\footnote{The term "Vainshtein screening" denotes a nonlinear effect that is at play in the DGP model and that dynamically suppresses the size of the modifications to gravity in regions where the enclosed matter density is large.} suppression. 

The second effect is that the distribution of halo masses differs between the DGP and the $\lcdm$ mock. Figure 3 of \cite{Barreira:2016a} shows that mock CMASS galaxies live in lower mass halos with increasing fifth force strength. These differences in the HOD models arise to preserve the galaxy number density and large-scale amplitude of the power spectrum monopole given modified halo abundances, halo bias and linear matter power spectrum. Because the DGP mocks contain more low mass halos, the amplitude of the lensing signal is suppressed relative to $\lcdm$. 
 
 Overall, Figure \ref{mgplot} shows that these two competing effects result in a difference to $\Delta\Sigma$ that is scale dependent, reflecting the regimes where each of these two effects dominate. Of the three DGP cases shown, the DGPs and DGPm ones are those which have the largest impact on $\Delta\Sigma$. However, these two particular models are already severely disfavored by current growth rate measurements \cite{Barreira:2016a}. The DGPw case has a goodness-of-fit to the growth rate data that is comparable to $\lcdm$, but its impact on $\Delta\Sigma$ does not exceed $5\%$, thereby falling short of the $30\%$ mismatch displayed in Figure \ref{bossmocks}. Furthermore, the DGP gravity models that we have explored predict a scale dependence in the lensing amplitude, which is inconsistent with our observations which suggest a fairly scale-independent offset. These tests suggest that the mismatch between the $\lcdm$ mocks and data in Figure \ref{bossmocks} is unlikely to be solely explained by DGP-like modifications to gravity, or other theories with similar phenomenology.

\begin{figure}
\begin{center}
\includegraphics[width=8.5cm]{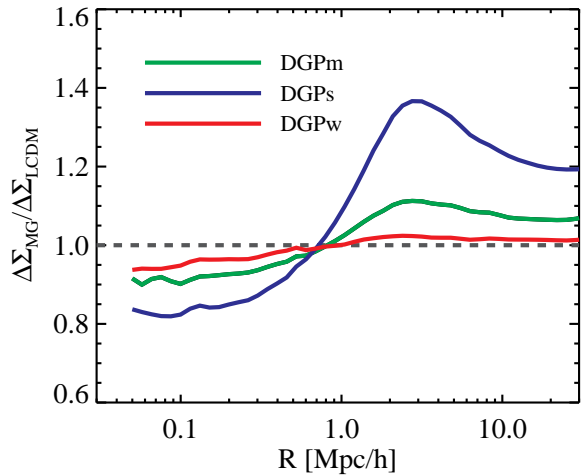}
\caption{Ratio of $\Delta\Sigma$ from DGP gravity simulations and $\Delta\Sigma$ from $\lcdm$. The three curves show the result for varying strength of the modified gravity effects, with DGPs, DGPm and DGPw corresponding to strong, medium and weak departures from GR, respectively. When the galaxy clustering strength is held fixed, departures from GR leave a scale-dependent signature in $\Delta\Sigma$.}
\label{mgplot}
\end{center}
\end{figure}


\section{Summary and Conclusions}\label{conclusions}

We report high signal-to-noise g-g lensing measurements ($S/N=30$) for the BOSS CMASS sample of massive galaxies at $z\sim0.55$ using 250 square degrees of weak lensing data from the CFHTLenS and CS82 surveys. We compare the amplitude of this signal with predictions from mock catalogs trained to match a variety of observables including the galaxy stellar mass function, the projected correlation function, and the two-dimensional redshift space clustering of CMASS. All models yield surprisingly similar prediction for the lensing observable $\Delta\Sigma$ with differences that are at the 20 percent level at most (with most models agreeing at the 15 per cent level). This is quite remarkable given significant differences in the methodologies (including both HOD and SHAM), cosmologies, and {\it N}-body simulations that were used to construct the models. We conclude that given standard assumptions about how
galaxies populate dark matter halos, the clustering of CMASS makes a
robust prediction for the amplitude of the lensing signal.

Figure \ref{bossmocks} corresponds to our main result which is the comparison between model predictions and the lensing measurement. This comparison reveals that the amplitude of the CMASS g-g lensing signal is 20-40\% lower than predicted from standard models of the galaxy-halo connection constrained by the clustering of CMASS. We present a detailed investigation of a range of systematic effects associated with lensing measurements, including the effects of \photoz errors, boost factors, and the effects of the lensing weight function. Our measurement is robust to all of these effects. Furthermore, our CS82 lensing catalog yields the same values for $\Delta\Sigma$ as an independent lensing measurements from SDSS. Our tests (the details of which are mostly given in the appendices)
show that the differences reported in Figure \ref{bossmocks} are too large to be explained by systematic effects alone and that the mismatch is a genuine effect.
 
This leads us to consider other explanations for the low lensing amplitude. The combination of g-g lensing and clustering is sensitive to \textcol{$S_{\rm 8}=\sigma_8 \sqrt{\Omega_m/0.3}$. We use an analytic HOD formalism to perform a joint fit to $\Delta\Sigma$ and to $w_p^{\rm CMASS}$ where $\sigma_8$ and $\Omega_{\rm m}$ are left as free parameters. Figure \ref{cosmologyplot} shows that lowering the value of $S_{\rm 8}$ by 2-3$\sigma$ compared to Planck 2015 reconciles the lensing with clustering.} Because our measurements are dominated by non-linear scales where the details of the galaxy-halo connection matter, these results alone should not be construed as evidence for a low value of $S_8$. However, the cosmological interpretation of these results does become more interesting when considered in the context of multiple constraints on the amplitude of low-redshift structure, both from lensing and from cluster abundances \citep[\eg][]{Heymans:2013,Planck-Collaboration:2015clusters, Hildebrandt:2016, Joudaki:2016}, that yield lower amplitudes compared to the Planck 2015 $\Lambda$CDM predictions (but also see \citealt[][]{Jee:2013}, \citealt{The-Dark-Energy-Survey-Collaboration:2015}, and \citealt[][]{Kitching:2016}).

The model predictions shown in Figure \ref{bossmocks} use standard galaxy-halo modeling based on either HOD or SHAM type methodologies. The fact that the amplitude of the lensing does not match the predictions from these models may reflect an inherent failure of such models. If the ages of galaxies correlate with the ages of their dark matter halos, then assembly bias effects may be present in color selected samples such as CMASS. The clustering of CMASS tightly constrains the large-scales bias of the sample whereas the lensing is sensitive to the mean halo mass. The difference that we observe may suggest a tension between the halo mass and the large-scale bias of this sample -- the smoking gun for assembly bias. If assembly bias is at play, it could be a systematic effect for RSD constraints from upcoming surveys such as DESI \citep{Levi:2013}. Lensing measurements such as presented in this paper can play an important role in understanding theoretical systematic uncertainties associated with the complexities of galaxy bias. 

Another effect that may be non-negligible given the precision of our measurements is the impact of baryon physics processes on the matter distribution. We use the Illustris simulations to present a first estimate of the magnitude of baryonic effects on the weak lensing profiles of subhalo-abundance matched galaxies at BOSS CMASS-like number densities and redshifts. We find that baryonic effects can induce a 10-30\% difference in $\Delta\Sigma$ (with a characteristic scale-dependence) and a factor of two difference in the satellite fraction for CMASS-like galaxy samples. This level of difference is no longer negligible given the statistical errors on our measurements. Without further analysis, it is difficult to say exactly how these effects would play out in an HOD or SHAM analysis of the clustering of BOSS galaxies, and whether or not the differences go in the same direction as our lensing measurements.

We also consider the impact of finite mass neutrinos on $\Delta\Sigma$. We run three $N$-body simulations with the same value for the total matter density but with different neutrino masses (0eV, 0.15eV and 0.3eV). We show that the effect of massive neutrinos goes in the right direction to explain the low amplitude of our lensing signal. However, the impact of massive neutrinos on $\Delta\Sigma$ is at the $\sim$10\% level at most, and so massive neutrinos alone are unlikely to be the full story.

Finally, we investigate the impact of modified gravity on $\Delta\Sigma$ and show that the existence of a positive additional ``fifth" force common to many modified gravity models leaves a scale dependent signature in the lensing signal. The amplitude of this effect, combined with the fact that our reported difference is fairly scale independent, leads us to conclude that modified gravity effects are unlikely to explain the difference reported in this paper. 
 
The mismatch that we report could be due to one, or a combination of the effects described above. Disentangling cosmological effects from the details of the galaxy-halo connection, the effects of baryons, and finite mass neutrinos, is the next challenge facing joint lensing and clustering analyses. This is especially true in the context of large galaxy samples from Baryon Acoustic Oscillation surveys with precise measurements but complex selection functions.

\section*{Acknowledgements}

We would like to thank Melanie Simet, Hironao Miyatake, Eric Jullo, Massimo Viola, Eduardo Rozo, and Andrew Hearin for many useful conversations during the preparation of this manuscript. This work is based on observations obtained with MegaPrime/MegaCam, a
joint project of CFHT and CEA/IRFU, at the Canada-France-Hawaii
Telescope (CFHT) which is operated by the National Research Council
(NRC) of Canada, the Institut National des Sciences de l'Univers of
the Centre National de la Recherche Scientifique (CNRS) of France, and
the University of Hawaii. This research used the facilities of the
Canadian Astronomy Data Centre operated by the National Research
Council of Canada with the support of the Canadian Space
Agency. CFHTLenS data processing was made possible thanks to
significant computing support from the NSERC Research Tools and
Instruments grant program. This work was supported by World Premier International Research Center Initiative (WPI), MEXT, Japan and JSPS KAKENHI Grant number JP15K17601. Funding for SDSS-III has been provided by the Alfred P. Sloan
Foundation, the Participating Institutions, the National Science
Foundation, and the U.S. Department of Energy Office of Science. The
SDSS-III web site is \url{http://www.sdss3.org/}. SDSS-III is managed
by the Astrophysical Research Consortium for the Participating
Institutions of the SDSS-III Collaboration including the University of
Arizona, the Brazilian Participation Group, Brookhaven National
Laboratory, Carnegie Mellon University, University of Florida, the
French Participation Group, the German Participation Group, Harvard
University, the Instituto de Astrofisica de Canarias, the Michigan
State/Notre Dame/JINA Participation Group, Johns Hopkins University,
Lawrence Berkeley National Laboratory, Max Planck Institute for
Astrophysics, Max Planck Institute for Extraterrestrial Physics, New
Mexico State University, New York University, Ohio State University,
Pennsylvania State University, University of Portsmouth, Princeton
University, the Spanish Participation Group, University of Tokyo,
University of Utah, Vanderbilt University, University of Virginia,
University of Washington, and Yale University. The MultiDark Database
used in this paper and the web application providing online access to
it were constructed as part of the activities of the German
Astrophysical Virtual Observatory as result of a collaboration
between the Leibniz-Institute for Astrophysics Potsdam (AIP) and the
Spanish MultiDark Consolider Project CSD2009- 00064. The Bolshoi and
MultiDark simulations were run on the NASA’s Pleiades supercomputer at
the NASA Ames Research Center. The MultiDark-Planck (MDPL) and the
BigMD simulation suite have been performed in the Supermuc
supercomputer at LRZ using time granted by PRACE. Numerical
computations were partly carried out on Cray XC30 at Center for
Computational Astrophysics, National Astronomical Observatory of
Japan. PB was supported by program number HST-HF2-51353.001-A, provided by NASA through a Hubble Fellowship grant from the Space Telescope Science Institute, which is operated by the Association of Universities for Research in Astronomy, Incorporated, under NASA contract NAS5-26555. TE is supported by the Deutsche
Forschungsgemeinschaft in the framework of the TR33 ‘The Dark
Universe’. SH acknowledges support by the DFG cluster of excellence \lq{}Origin and Structure of the Universe\rq{} (\href{http://www.universe-cluster.de}{\texttt{www.universe-cluster.de}}). HYS acknowledges the support from Marie-Curie International Incoming Fellowship (FP7-PEOPLE-2012-IIF/327561) and NSFC of China under grants 11103011. CH was supported by the European Research Council under grant number 647112. HH is supported by an Emmy Noether grant (No. HI 1495/2-1) of the Deutsche Forschungsgemeinschaft. MV and FVN are supported by ERC-StG "cosmoIGM" and PD51 Indark Grant. FS acknowledges support from the Marie Curie Career Integration Grant(FP7-PEOPLE-2013-CIG) ``FundPhysicsAndLSS". RM acknowledges the support of the Department of Energy Early Career Award program.


\appendix

\section{Cross-Checks and Weak Lensing Systematic Tests}

\subsection{On the Computation of a Boost Correction Factor to
  account for Physically Associated Galaxies}\label{appendix_boost}

As described in Section \ref{backgroundselection}, we use \photoz cuts
to select background source galaxies ($z_S>z_L$). However, because
\photoz estimates are far from perfect, our ``background'' sample may
contain a number of galaxies are either actually in the foreground
($z_S<z_L$), or which are physically associated with the lens sample
($z_S=z_L$). Because foreground and physically associated galaxies are
unlensed, the inclusion of these galaxies will cause $\Delta\Sigma$ to
be underestimated (``dilution'' effect). The exact magnitude of this
effect will depend on the quality of the photo-$z$s, as well as the
details of the lens-source separation cuts.

A “boost correction factor” is sometimes applied in order to account for
the dilution of the signal by physically associated sources 
\citep[\eg][]{Kneib:2003aa,Sheldon:2004, Hirata:2004aa,
  Mandelbaum:2006}. This correction factor is usually computed by
comparing the weighted number density of source galaxies for the lens
sample to the weighted number density of source galaxies around random
points:

\begin{equation}
C(r) = \frac{N_{\rm rand}}{N_{\rm lens}}\frac{\sum_{\rm lens} w_{\rm
    lens}}{\sum_{\rm rand} w_{\rm rand}}
\end{equation}

However, a key assumption underlying this procedure is that physically
associated galaxies are the dominant contribution to $C(r)$. In
practice, other effects may also modify the number density of source
galaxies as a function of lens-centric distance such as magnification
bias \citep[][]{Mandelbaum:2006}, obscuration effects
\citep{Melchior:2015aa,Applegate:2014aa,Simet:2015}, and local
galaxy dependent quality cuts
\citep{Melchior:2015aa,Applegate:2014aa}. The later is particularly
pernicious for the CFHTLenS and CS82 catalogs due to conservative
deblending settings used by the {\it lens}fit shape measurement
algorithm. Another effect which has been less discussed, is the
availability of a photometric redshift. Indeed, in addition to shape
measurements, photometry measurements may also be more likely to fail in
high density regions which would impact the radial density profile of
galaxies with reliable photo-$z$s. If these effects are not taken into account, boost-factors will be mis-estimated.

To illustrate the impact of lensing and \photoz quality cuts on the
radial source density profile, we compute the number of source
galaxies in the CS82 catalog as a function of lens-centric distance
after applying each of the following cuts in order:

\begin{enumerate}
\item Remove objects classified as stars by {\it lens}fit ({\sc
    fitclass}$=1$) as well as objects in masked regions.
\item Remove blended objects ({\sc fitclass}$=-2$).
\item Apply a {\sc fitclass}$=0$ cut. This cut removes objects
  which have a bad fit, or for which the chi-squared exceeds a critical
  value.
\item Select galaxies with a non zero lensing weight ($w>0$).
\item Select galaxies with $z_{\rm phot}>0$ and {\sc odds}$>0.5$.
\end{enumerate}

Figure \ref{boost_factors} displays the results of this exercise and
demonstrates that lensing and \photoz quality cuts have a non trivial
impact on the radial density profile of source galaxies out to scales
of at least 1 $h^{-1}$ Mpc. At first glance, it may tempting to think
that Figure \ref{boost_factors} provides a straightforward
characterization of the impact of each of these cuts. However, it is
important to remember that each cut removes a set of legitimate
background galaxies, but also modifies the number of physically
associated pairs (there is no reason this number should remain
constant after each cut) -- and disentangling these two effects is non
trivial. The best approach so far to this problem has been to
characterize the combined effects of obscuration and of the lensing
quality cuts by computing the recovery rate of fake galaxies inserted
into real images \citep{Melchior:2015aa}. However, in addition to these effects, \photoz quality cuts may also have a non
trivial local galaxy dependance. This effect has been less discussed so
far but warrants further attention.

Given these difficulties, we do not apply boost
correction factors in this paper. Instead, we adopt a more empirical
approach and check that our lensing signal is robust to various lens-source separation cuts. Indeed, if our signal
is affected by a dilution effect, then we should find that the
amplitude of the lensing signal on small scales increases as we implement more conservative lens-source separation cuts. In the
following section, we demonstrate that we do not observe this effect
-- suggesting that our lens-source separation cuts are stringent
enough that our lensing signals do not suffer from a dilution caused
by physically associated galaxies. However, it is clear that these effects warrant closer scrutiny using simulations such as presented by \citet{Melchior:2015aa}.

\begin{figure*}
\begin{center}
\includegraphics[width=17cm]{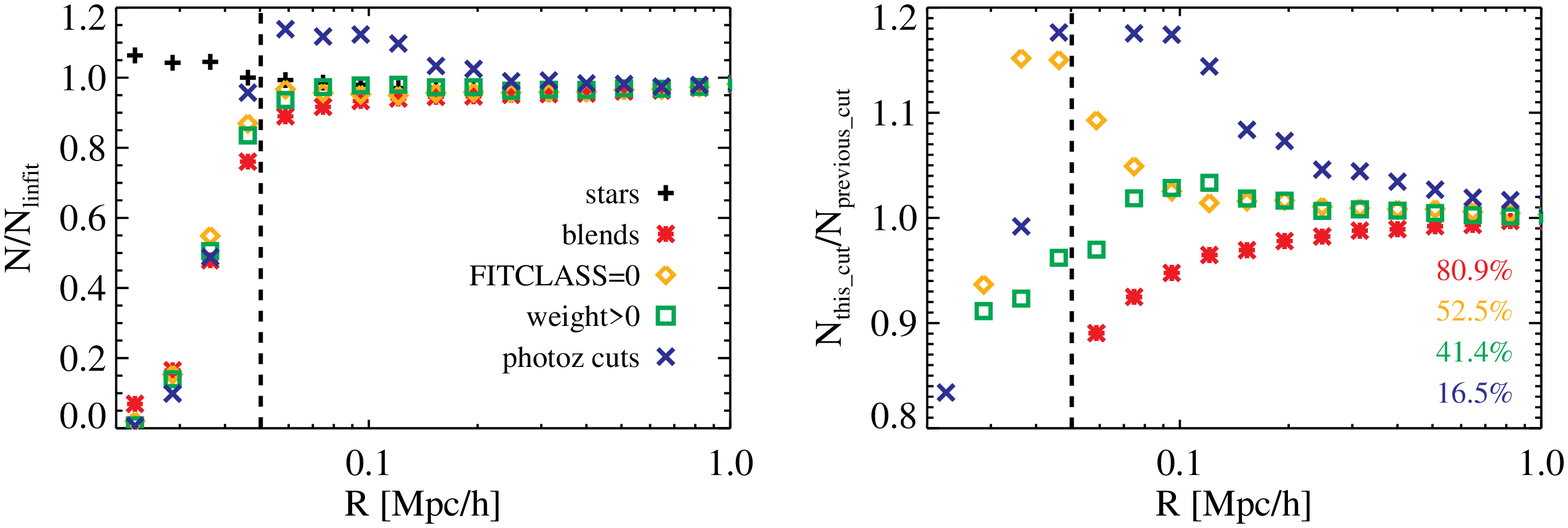}
\caption{Number of source galaxies in the CS82 catalog as a function
  of lens-centric distance after applying each of our lensing \photoz
  quality cuts. In this exercise, all source galaxies are artificially
  placed at $z=0.8$. Left: number of objects in the source catalog
  after each cut, divided by the expected value from a linear fit to
  the number of objects at $r>1$ \h Mpc. The dashed vertical line
  indicates the minimum radial scale of our lensing
  measurement. Right: number of objects in the source catalog divided
  by the number from the previous cut. Numbers in the right hand panel
  indicate the fraction of galaxies that remain in the catalog after
  each cut compared to the initial number. Lensing and \photoz quality
  cuts impact the radial source density profile out to scales of at
  least 1 $h^{-1}$ Mpc.}
\label{boost_factors}
\end{center}
\end{figure*}

\subsection{Photometric Redshifts}\label{appendix_photoz}

In this section, we present a series of tests to verify that our
lensing signals are robust to a variety of different photometric
redshift cuts. No statistically significant systematic trends are found for any of the tests
that we have implemented.

First, we show that our lensing signal is robust with respect to {\sc BPZ}
{\sc odds} parameter cuts. Figure \ref{zphot_tests} presents the CMASS
lensing signal computed for three different odds cuts ({\sc odds}$>0$,
{\sc odds}$>0.4$, and {\sc odds}$>0.8$). The fact that the amplitude
of the signal is insensitive to this {\sc odds} cut suggests that our
signal is relatively robust to systematic errors due to catastrophic
outliers.

Second, \citet{Hildebrandt:2012} and \citet{Benjamin:2013aa} caution
that the quality of the CFHTLenS photometric redshifts degrade at
$z_S>1.3$. Our fiducial source catalog does not include a high
redshift cut. To test if this choice impacts our results, we compute
the lensing signal using only source galaxies with $z_S<1.3$ and show
the results in Figure \ref{zphot_tests}. We find no statistically
significant shift in the signal when we enforce a source redshift cut
at $z_S<1.3$. A similar test with consistent results for CFHTLenS is presented in Figure C2 of \citet[][]{Coupon:2015}.

Third, we test if our results are robust with respect to the
lens-source separation cuts. We consider three different schemes for
isolating background galaxies:

\begin{itemize}
\item {\sc zcut1} : $z_{S}>z_{L}+0.1$ and $z_{S}>z_{L}+\sigma_{95}$
\item {\sc zcut2} : $z_{S}>z_{L}+0.1$ and $z_{S}>z_{L}+\sigma_{95}/2.0$
\item {\sc zcut3} : $z_{S}>z_{L}+0.1$
\end{itemize}

Here, {\sc zcut1} is a more conservative choice than {\sc zcut3}. Our
fiducial lens-source separation cut is {\sc zcut2}. We compute the
CMASS lensing signal using each of these three lens-source separation
schemes and display the results in Figure \ref{zphot_tests}. The
amplitude of our lensing signal does not vary when we enforce a more
stringent lens-source separation scheme which suggests that our
lensing signals do not suffer from a dilution caused by physically
associated galaxies.

\begin{figure*}
\begin{center}
\includegraphics[width=16cm]{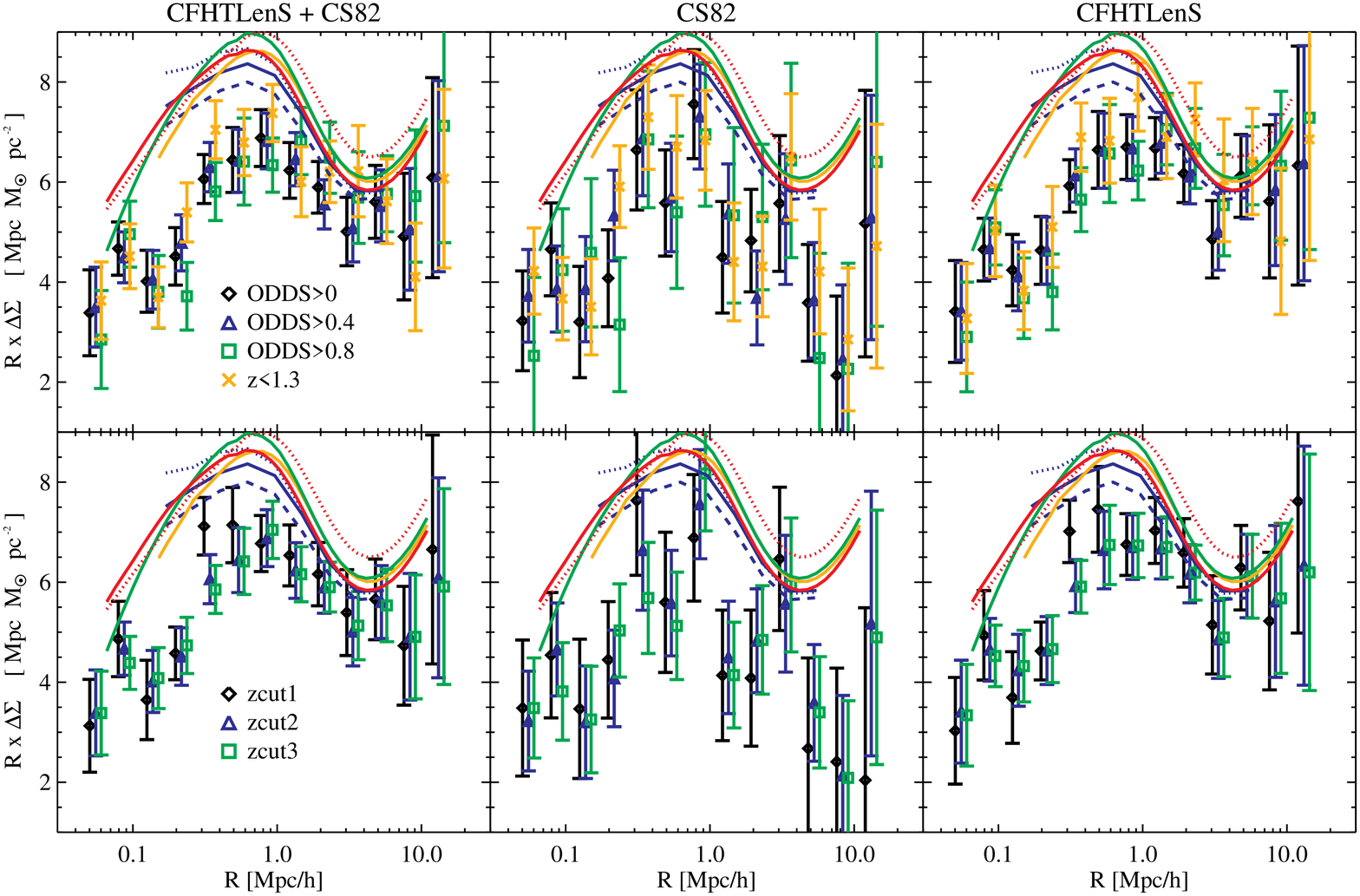}
\caption{Tests for systematic effects associated with photometric
  redshifts. Left panels: combined CS82 and CFHTLenS lensing
  signal. Middle panels: CS82 lensing signal. Right panels: CFHTLenS
  lensing signal. Upper panels show the lensing signal computed for
  several different cuts on the {\sc odds} parameter and when the
  source sample is restricted to $z_S<1.3$. We do not display the results for $z_S>1.3$ simply because only a small fraction of the source sample lies at $z_S>1.3$ and the signal becomes too noisy to make a useful comparison. Lower panels show the
  lensing signal computed for three different lens-source separation
  cuts. As described in Section \ref{appendix_photoz}, {\sc zcut1} is a more
  conservative choice than {\sc zcut3} for selecting background
  galaxies. No statistically significant systematic trends are found
  for any of these tests. Lines represent model predictions using the same color scheme as in Figure \ref{bossmocks}. The CFHTLenS measurements appear to be more consistent with the model predictions at $r>2$ \h Mpc than CS82. However, as argued in Section \ref{errors}, there is significant field-to-field variance on these scales which means that the combined CS82+CFHTLenS measurement (left panels) should be more robust than either measurement alone.}
\label{zphot_tests}
\end{center}
\end{figure*}


Finally, we use the combined spectroscopic redshift catalog described
in Section \ref{photoz} to estimate the level of \photoz bias in
$\Delta\Sigma$ for CS82. To correct for spectroscopic incompleteness and to ensure that the spectroscopic sample has the same distribution as our source sample, we use the weighting scheme described in \citet[][]{Hildebrandt:2016} which follow ideas originally outlined by \citet{Lima:2008}. This method determines the density in five-dimensional magnitude space of spectroscopic objects as well as objects in the lensing catalogue via a k-nearest neighbor estimate. The ratio of the densities at the position of each spectroscopic object is then used as a weight for this particular object. There are two main requirements for this method which are: a) the spectroscopic catalogue must cover the whole extent of the lensing catalog in magnitude space and b) the mapping from magnitudes to redshifts must be unique. In our case, the first condition is satisfied. Even before re-weighting, the distribution of spec-$z$ objects in magnitude space is very similar to the distribution of our source sample. Hence the weights are rather small and only a mild re-weighting is necessary. The second requirement, however, is more difficult to quantify \cite[see][]{Lima:2008,Cunha:2009, Cunha:2012, Cunha:2014}. However, given that we are using mostly objects with $i<24$, which do not extend to very high redshifts, strong degeneracies are not expected. 

We now outline our procedure to estimate the bias on $\Delta\Sigma$ arising from photo-$z$s by using our re-weighted spectroscopic sample \citep[][]{Mandelbaum:2008,Nakajima:2012aa}. Let $\Delta\Sigma_{\rm P}$ represent the (possibly biased) value of $\Delta\Sigma$ measured with \photozs, and let $\Delta\Sigma_{\rm T}$ represent the true value of $\Delta\Sigma$. Likewise, let $\Sigma_{\mathrm{crit,P}}$ represent the value computed from photo-$z$s and $\Sigma_{\mathrm{crit,T}}$ represent the true value of $\Sigma_{\mathrm{crit}}$. The true value of the gravitational shear is simply:

\begin{equation}
\gamma_{\rm T}=\Delta\Sigma_{\rm T}/\Sigma_{\mathrm{crit,T}}
\label{trueshear}
\end{equation}

If a source is at $z_S<z_L$, then $\gamma_{\rm T}=0$ (this accounts for the dilution effect by sources that scatter above $z_L$ but which are actually located at lower redshifts than $z_L$).  By combining Equation \ref{trueshear} with Equation \ref{ds_equation}, we can form as estimate of $f_{\rm bias}=\Delta\Sigma_{\rm T}/\Delta\Sigma_{\rm P}$ via:

\begin{equation}
f_{\rm bias}^{-1} = 
{\sum_{i=1}^{N_{S}} w_{{\rm spec},i} \left( \Sigma_{{\rm crit,P},i}/\Sigma_{{\rm crit,T},i} \right)
 \over \sum_{i=1}^{N_{S}}w_{{\rm spec},i}} 
\label{ds_equation3}
\end{equation}

\noindent where the sum is performed over all source galaxies with spectroscopic redshifts and where the weight $w_{\rm spec}$ is analogous to $w_{\rm ds}$ but include the additional spectroscopic redshift weight described previously. When computing Equation \ref{ds_equation3} we randomly draw redshifts from our lens sample. Via this procedure, we find $f_{\rm bias}=0.97$ which suggests that biases due to photo-$z$s are at the 3\% level and are not a concern for this work. This estimate includes the dilution of the signal by source galaxies with $z_{S}^{\rm spec}<z_L$ but $z_{S}^{\rm phot}>z_L$.

\subsection{Signal Around Random Points}\label{appendix_randoms}

As a test for systematic effects, we also compute the stacked lensing
signal around a set of random points drawn from the same redshift
distribution as our CMASS lens sample. For each of the two surveys,
the density of random points is set to 100 times the density of the
CMASS sample and errors on the signal around random points are
computed via bootstrap. The result are presented in Figure
\ref{rand_tests}. No statistically significant systematic shear
patterns are detected around random points for either $\Delta\Sigma_{\rm t}$
or $\Delta\Sigma_{45}$. We note that with this density of random points, the signal around random points becomes highly correlated on large-scales due to correlated shape noise. 

\begin{figure*}
\begin{center}
\includegraphics[width=16cm]{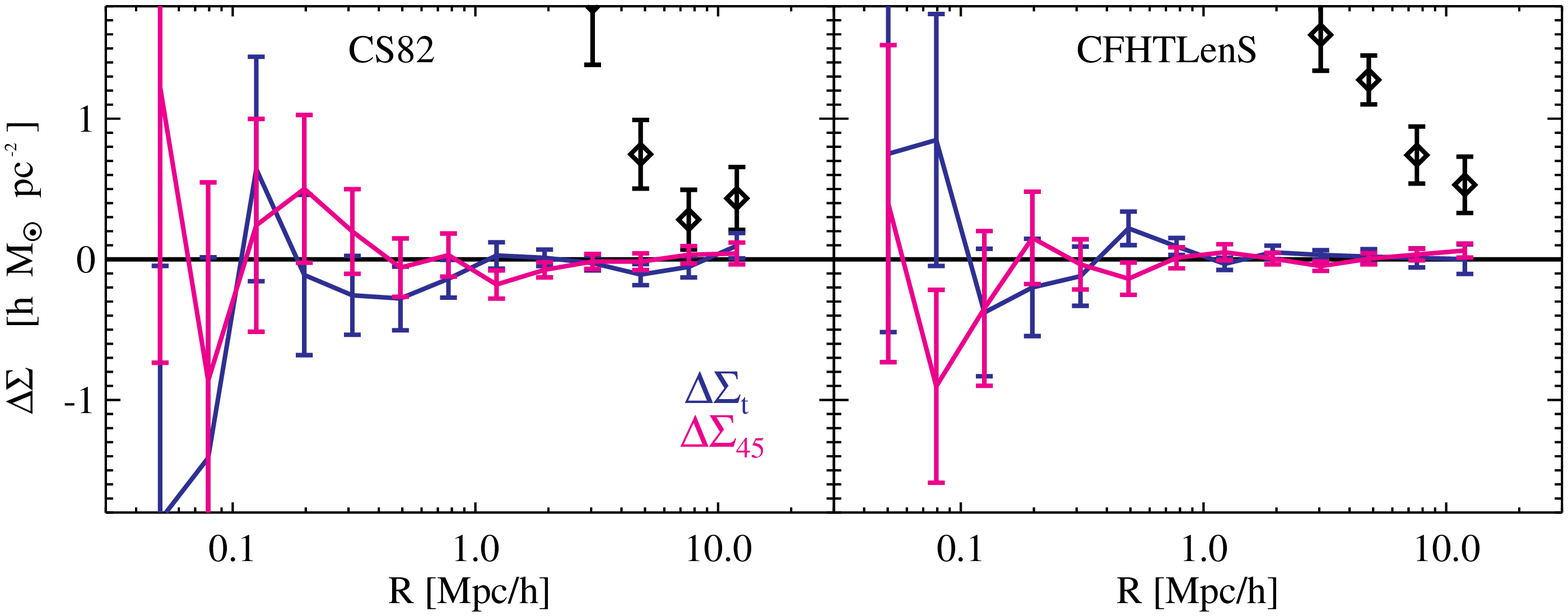}
\caption{Stacked lensing signal around random points for CS82 (left)
  and CFHTLenS (right). Errors are computed via bootstrap. Black
  diamonds show the stacked lensing signal for CMASS galaxies. No
  statistically significant systematic shear patterns are detected
  around random points for either $\Delta\Sigma_{\rm t}$ or
  $\Delta\Sigma_{45}$. The signal around random points becomes highly correlated on large-scales due to correlated shape noise.}
\label{rand_tests}
\end{center}
\end{figure*}

\subsection{Fiber Collisions and Redshift Failures}\label{app:fibercollisions}

As discussed in Section \ref{bosssample}, a small number of galaxies
from the CMASS target catalog do not have a spectroscopic redshift
because of fiber-collision effects and redshift measurement
failures. We test four different schemes designed to account for these
missing galaxies:

\begin{itemize}
\item {\sc wht} : the nearest neighbor based weighting scheme adopted in
  \citet{Anderson:2012}.  In this approach, only galaxies which have a
  measured spectroscopic redshift are used when computing the lensing
  signal. Galaxies have a weight equal to $w_{\rm tot}=w_{\rm rf} +
w_{\rm fc}-1 $.
\item {\sc nn} : galaxies which do not have a redshift are assigned
  the same redshift as their nearest neighbor ($z_{\rm NN}$). In
  contrast with the previous method, galaxies which do not have a
  spectroscopic redshift \emph{are} used when computing the lensing
  signal (with a redshift set to $z_{\rm NN}$).
\item {\sc zphot} : CMASS galaxies which do not have a spectroscopic
  redshift are assigned a \photoz using the CFHTLenS and
  CS82 \photoz catalogs.
\item {\sc discard} : galaxies which do not have a spectroscopic redshift
  are removed from the catalog. No additional weighting scheme is
  applied. 
\end{itemize}

Figure \ref{fiber_collisions} demonstrates that the measured lensing
does not depend strongly on the correction scheme for galaxies which
do not have a spectroscopic redshift. The right hand panel of Figure
\ref{fiber_collisions} shows the impact of ignoring this effect
altogether (the {\sc discard} scheme). The impact of missing redshifts
from fiber collisions and redshift failures is small, but ignoring
this effect altogether may lead to a small underestimate of the CMASS
lensing signal because fiber collisions tend to occur in high density
regions (see discussion in \citealt{Reid:2014}). For our fiducial
signal, we adopt the nearest neighbor based weighting scheme ({\sc
  wht}).

\begin{figure*}
\begin{center}
\includegraphics[width=15cm]{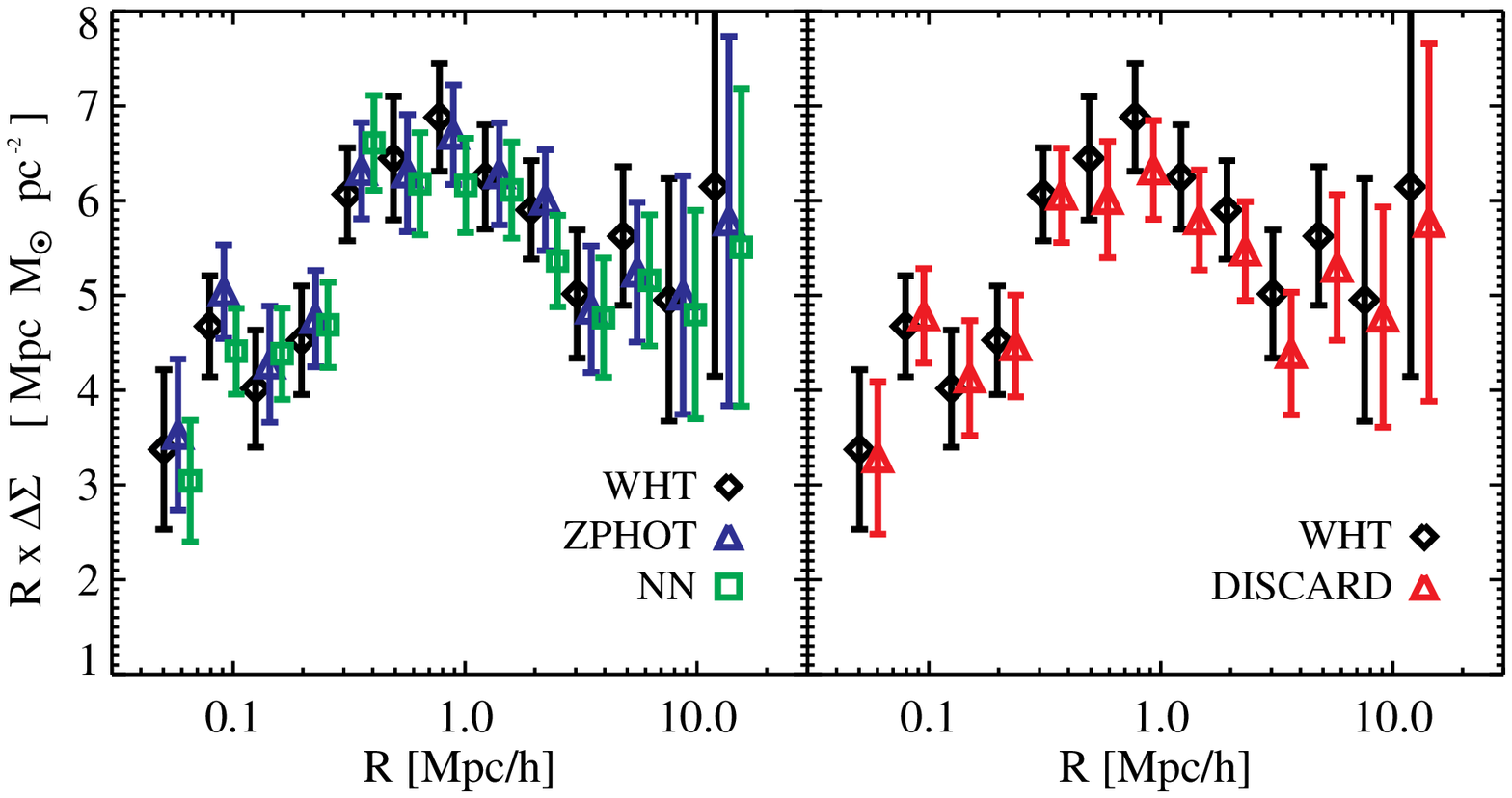}
\caption{{\it Left}: lensing signal computed using three different
  methods to account for CMASS galaxies with missing spectroscopic
  redshifts. All three schemes yield similar lensing signals. {\it
    Right}: impact of ignoring objects with missing redshifts ({\sc
    discard}, red triangles) compared to our fiducial nearest
  neighbor based weighting scheme ({\sc wht}, black triangles). The
  red data points are systematically lower than the black data points
  suggesting that ignoring this effect altogether may lead to a small
  underestimate of the CMASS lensing signal because fiber collisions
  tend to occur in high density regions \citep[][]{Reid:2014}.}
\label{fiber_collisions}
\end{center}
\end{figure*}

\subsection{Weighting of the Lensing Signal}\label{app:weighting}

By examining Equation \ref{ds_equation}, we see that not all lens galaxies will contribute an equal weight to $\Delta\Sigma$ \citep[also see, ][]{Nakajima:2012aa,Mandelbaum:2013,Simet:2016}. To highlight this, let us consider a single lens, $i$, and adopt the notations:

\begin{equation}
\Delta\Sigma_i=
 {\sum_{j=1}^{N_{S}} w_{{\rm ds},ij} \times \gamma_{{\rm t},ij}\times \Sigma_{{\rm crit},ij}
    \over \sum_{j=1}^{N_{S}}w_{{\rm ds},ij}} 
\end{equation}

\begin{equation}
w_{\rm{lens},i}=\sum_{j=1}^{N_{S}}w_{{\rm ds},ij}
\end{equation}

\noindent Using this notation, Equation \ref{ds_equation} can be re-written as:

\begin{equation}
\Delta\Sigma = { \sum_{i=1}^{N_{L}} w_{\rm{lens},i} \times \Delta\Sigma_i \over \sum_{i=1}^{N_{L}} w_{\rm{lens},i}}
\end{equation}

\noindent This is simply stating that in each radial bin, each lens is contributing to $\Delta\Sigma$ with a weight given by $w_{\rm{lens},i}$. There are several reasons why $w_{\rm{lens},i}$ will differ from lens-to-lens:

\begin{enumerate}
\item There is a simple geometric effect in which the number of source galaxies per bin is redshift dependent when the bin size is fixed in comoving (or physical) units. 
\item Lenses at higher redshifts will have fewer source galaxies behind them. 
\item Lenses at higher redshifts will have a lower lensing efficiency (this is the $\Sigma_\mathrm{crit}^{-2}$ term in $w_\mathrm{ds}$). 
\item Sources at higher redshifts will have a larger shape measurement uncertainty (this is the $w$ term in $w_\mathrm{ds}$). 
\item Obscuration and deblending effects mean that we loose a certain fraction of source galaxies on small radial scales \citep[see Figure \ref{boost_factors} and discussion in][]{Simet:2015}. Because we expect obscuration and deblending effects to be more important for the central galaxies of massive halos, this could lead to both a radial and a halo mass dependence of $w_{\rm{lens},i}$ which would go in the direction of down-weighting the lensing signal from massive halos on small radial scales.
\end{enumerate}

Can we explain our results simply due to differences in the weight function for CMASS galaxies between lensing and clustering measurements? Because the lensing signal does not vary with redshift (see Figure \ref{combinedsignalfigure}), effects (i)-(iv) should have no impact on the lensing signal and so the main effect that we are concerned about is effect (v). \citet{Simet:2016} tackle the first four effects by computing an average per-lens weight that is applied directly to their models. However, this approach is more difficult to apply in the context of obscuration and deblending effects because it would require characterizing $w_{\rm{lens},i}$ as a function of halo mass. For this reason, we propose a simple empirical test that accounts for the first four effects and partially accounts for the fifth effect. Instead of stacking the lensing signal over all lens galaxies, we first compute $\Delta\Sigma_i$ individually for each CMASS lens. We then compute an unweighted version of $\Delta\Sigma$ by simply taking the average value of $\Delta\Sigma_i$ in each radial bin as: 

\begin{equation}
\Delta\Sigma_{\rm noweight} = { \sum_{i=1}^{N_{L}} \times \Delta\Sigma_i \over N_{L}}
\end{equation}

 In this stack, lenses are no longer weighted by $w_{\rm{lens},i}$. However, this procedure only partially accounts for obscurations because halos will still be down-weighted if obscuration effects are so large that there are no source galaxies in a given radial bin. Also, this estimator will have an increased variance compared to the traditional procedure because each lens is put on an equal footing instead of stacking by inverse variance.

\textcol{Figure \ref{reweighted} compares our fiducial signal with the reweighted signal computed following the procedure above. We find no evidence for a difference between our fiducial signal and the reweighted signal confirming our initial proposition that effects (i)-(iv) do not impact our lensing signal, but also suggesting that effect (v) is not large enough to be of concern. Finally, we remark that clustering measurements also have a specific weight function (because clustering measures pairs of galaxies, see for example \citealt{Mandelbaum:2011}). However, because the lensing is invariant with redshift, applying an extra redshift-dependent weight to put the lensing on the same footing as the clustering should also have no effect.}

\begin{figure}
\begin{center}
\includegraphics[width=8.5cm]{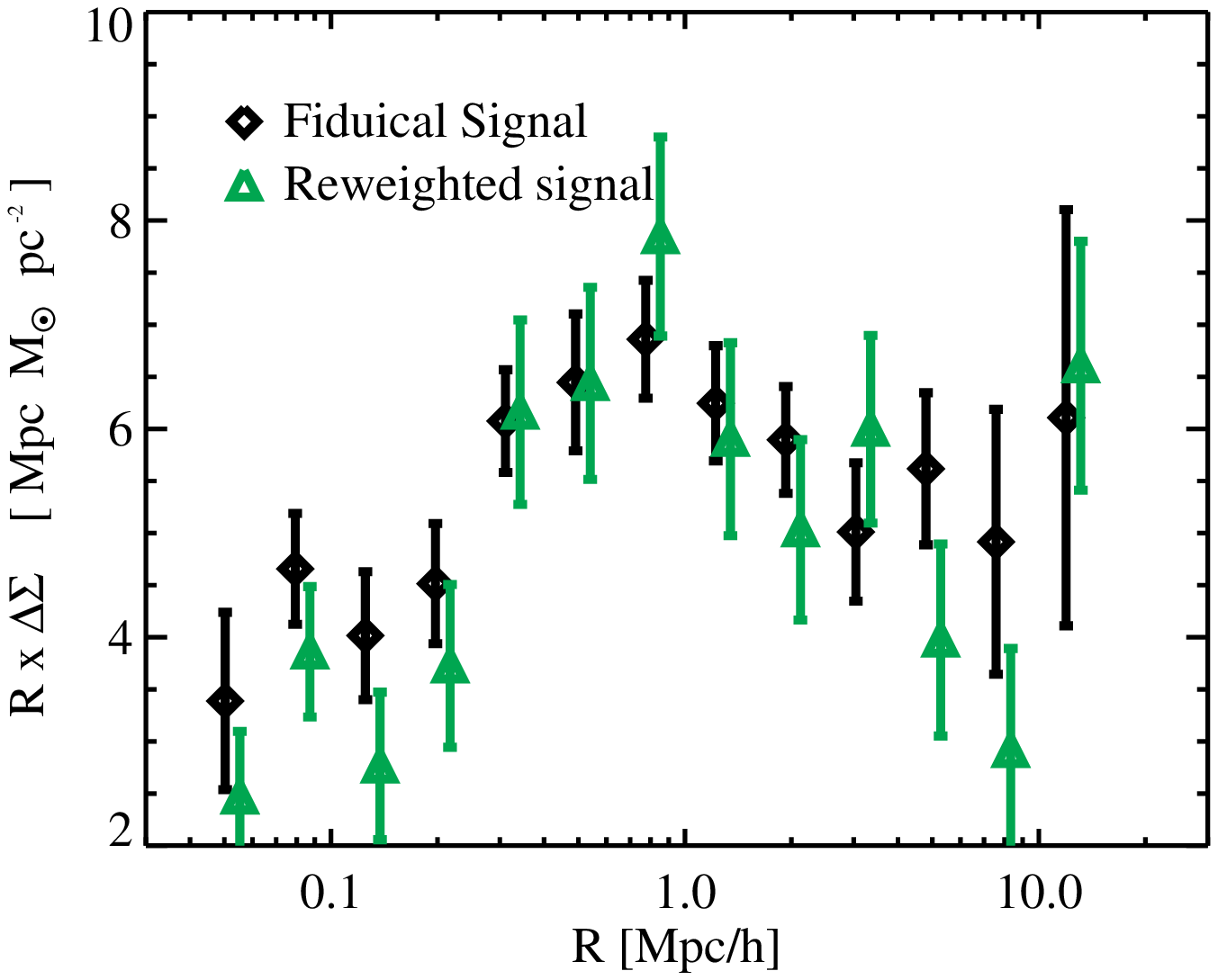}
\caption{Fiducial signal (black triangles) compared to reweighted signal (green triangles).}
\label{reweighted}
\end{center}
\end{figure}

\subsection{Comparison with Lensing from SDSS}\label{app:compsdss}

Here, we perform a cross-check on our CS82 lensing catalog by comparing with lensing from the SDSS catalog of \citet[][]{Reyes:2012aa} with correction factors for \photoz errors as derived by \citet{Nakajima:2012aa} and with the shear calibration described in \citet{Mandelbaum:2013}. We select a set of clusters from the redMaPPer cluster catalog \citep[v5.10, ][]{Rykoff:2014,Rozo:2014} with $0.1<z<0.3$ and with $\lambda>20$ where $\lambda$ is the cluster richness and we compute the lensing signal around this sample for both CS82 and SDSS. Figure \ref{compare_sdss} shows that the CS82 and SDSS lensing signals are in excellent agreement. This provides an overall sanity check on the CS82 lensing catalog including the combined effects of shear calibration bias and the quality of the photoz-$z$s. \textcol{The mean inverse-variance weighted offset between the two signals for 0.1-10 \h Mpc is (2.7$\pm$7.0)\%.}

\begin{figure}
\begin{center}
\includegraphics[width=8.5cm]{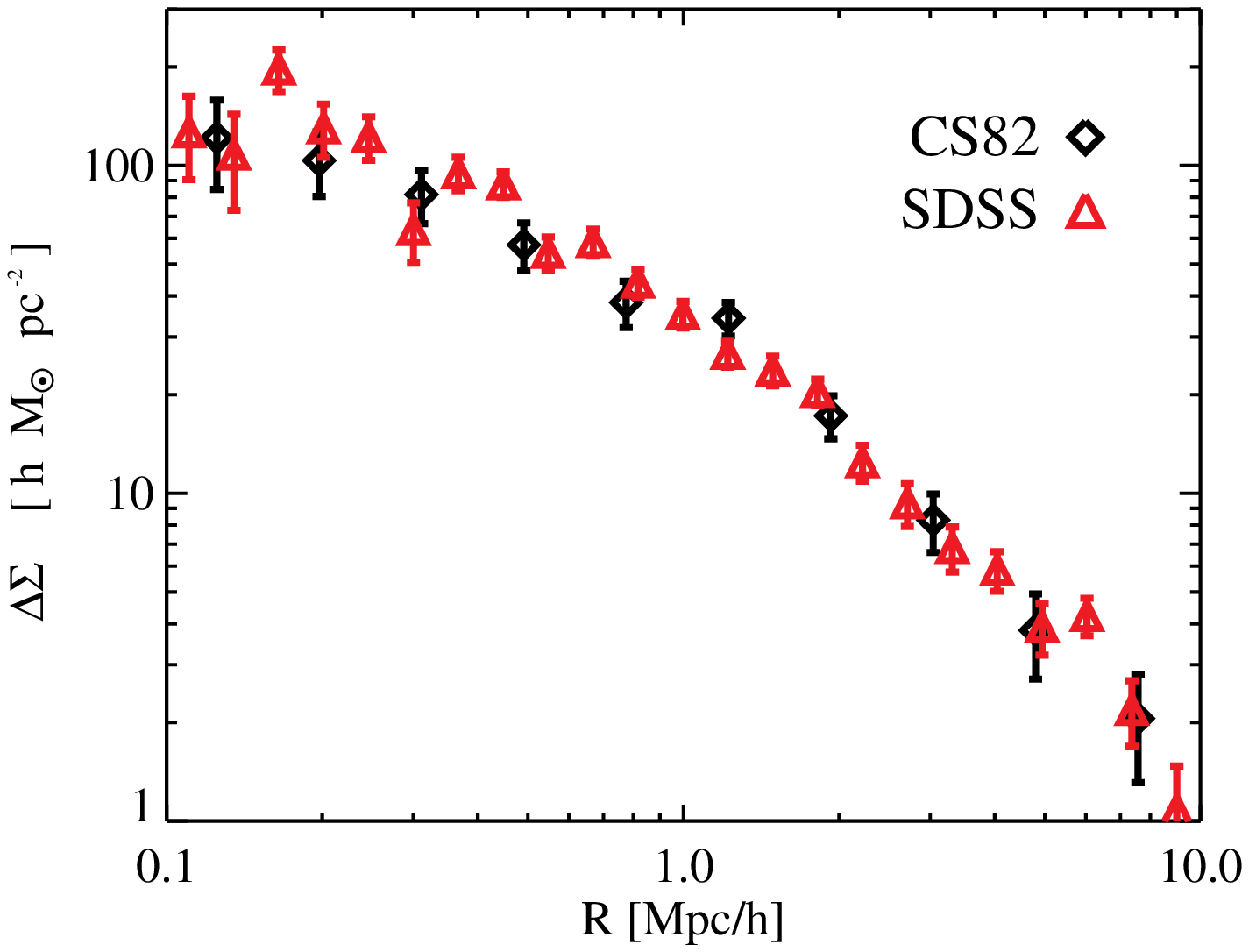}
\caption{Comparison between the lensing signal computed from SDSS and the lensing signal computed from CS82 for redMaPPer clusters with $0.1<z<0.3$ and with $\lambda>20$. The mean inverse-variance weighted offset between the two signals is consistent with zero.}
\label{compare_sdss}
\end{center}
\end{figure}

\subsection{Analytic HOD fit to Projected Correlation Function}\label{app:hodfit}

To construct Figure \ref{abtest}, we fit a simple four parameter analytic HOD model to the $w_{p}^{\rm CMASS}$ measurements of \citet[][]{Reid:2014} assuming a redshift of $z=0.55$. Our analytic HOD formalism is based on code described in previous work \citep[\eg][]{Leauthaud:2011,Leauthaud:2012a, Tinker:2013} and assumes the halo mass function calibration of \citet[][]{Tinker:2008} and the halo bias calibration of \citet[][]{Tinker:2010}. For consistency with our lensing measurements, we first convert the  \citet[][]{Reid:2014} measurements to a cosmology with $\Omega_{\rm m}=0.31$ following the methods outlined in \citet[][]{More:2013aa}. Our fit further assumes 
$\sigma_{8}=0.82$. Our HOD model assumes the following functional forms for the central and satellite occupation functions:

\begin{equation}
\langle N_{\rm cen}\rangle = \frac{1}{2}\left[ 1+\mbox{erf}\left(\frac{\log_{10}(M_{\rm h}) - \log_{10}(M_{\rm min}) }{\sigma_{\rm log M}} \right)\right]
\label{ncen}
\end{equation}

\begin{equation}
\langle N_{\rm sat} \rangle  = \langle N_{\rm cen} \rangle \left( \frac{M_h}{M_{\rm 1}}\right) ^{\alpha_{\rm sat}}
\end{equation}

\noindent where $M_{\rm h}$ is the halo mass defined as the mass enclosed by $R_{200b}$, the radius at which the mean interior density is
equal to 200 times the mean matter density. The four parameters varied in our fit are: $\log_{10}(M_{\rm min})$, $\log_{10}(M_{1})$, $\alpha_{\rm sat}$, $\sigma_{\rm log M}$. We set a broad prior such that  $3.2\times10^{-4} (h^{-1}\mathrm{Mpc})^{-3}< \overline{n} < 3.8\times10^{-4} (h^{-1}\mathrm{Mpc})^{-3}$. The best fit values from our fit are: $\log_{10}(M_{\rm min})=13.15 \pm 0.04$, $\log_{10}(M_{1})=14.26 \pm 0.05  $, $\alpha_{\rm sat}= 1.07 \pm 0.09$, and $\sigma_{\rm log M}=0.4540 \pm 0.06$.


\bibliographystyle{mnras}
\bibliography{mn-jour,all_refs}

\label{lastpage}


\end{document}